\documentclass[11pt]{article}
\usepackage{preamble} %

\title{\Large \textbf{Privacy Violations in Election Results}}

\author{Shiro Kuriwaki\thanks{Assistant Professor of Political Science and Resident Fellow at the Institution of Social and Policy Studies, Yale University.} \and Jeffrey B. Lewis\thanks{Professor of Political Science, University of California Los Angeles.} \and Michael Morse\thanks{Assistant Professor of Law and Political Science (by courtesy), University of Pennsylvania.}}

\date{\normalsize March 2025, \emph{Science Advances} (\textsc{doi}: \href{https://www.science.org/doi/10.1126/sciadv.adt1512}{\texttt{10.1126/sciadv.adt1512})}}

\begin{document}

\newgeometry{margin=1.3in}

\maketitle
\begin{abstract}
\noindent After an election, should election officials release a copy of each anonymous ballot? Some policymakers have championed public disclosure to counter distrust, but others worry that it might undermine ballot secrecy. 
We introduce the term \textit{vote revelation} to refer to the linkage of a vote on an anonymous ballot to the voter's name in the public voter file, and detail how such revelation could theoretically occur. 
Using the 2020 election in Maricopa County, Arizona, as a case study, we show that the release of individual ballot records would lead to no revelation of any vote choice for 99.83\% of voters as compared to 99.95\% under Maricopa's current practice of reporting aggregate results by precinct and method of voting. 
Further, revelation is overwhelmingly concentrated among the few voters who cast provisional ballots or federal-only ballots.
We discuss the potential benefits of transparency, compare remedies to reduce or eliminate privacy violations, and highlight the privacy-transparency tradeoff inherent in all election reporting.
\end{abstract}

\bigskip

\noindent \textbf{Teaser:} Voters' vote choices are rarely deducible even if cast vote records, indicating how every ballot cast was marked, are made public.

\newpage

\restoregeometry %
\frenchspacing %

\section{Introduction}

Democracies face a difficult tradeoff in reporting election results between promoting public transparency and protecting individual privacy. After votes are cast, election officials must report results at a level of granularity sufficient to bolster the legitimacy of the count while avoiding revealing individual choices. Where to strike that balance between transparency and privacy is an increasingly pressing question. That is because most modern voting systems can now preserve anonymous, individual ballot records in addition to tallying aggregate results. Further, at least since the 2020 election, a growing swath of voters have been seeking ``citizen audits'' of elections by making public records requests for these ballot records \parencite{stewart2022trust,pildes2021election,brennan2022_citizen}.

In theory, election results should both promote transparency and protect privacy.
Transparent, or granular, election results can promote confidence by helping to detect or deter electoral manipulation \parencite{wand2001butterfly, herron2019mail, goel2020one, cantu2019fingerprints, kronick2024}. 
Further, privacy, in the form of a secret ballot, can also reduce the market for vote buying and voter intimidation \parencite{keyssar2009right, mares2015open}. 
In practice, however, more transparent election results might unravel the secret ballot.

While privacy concerns with administrative data are not new \parencite{ohm2009broken,sweeney2017re,wood2018differential}, neither formal privacy scholars nor political scientists have systematically explored the particular privacy risks in election results. 
Without the benefit of theoretical or empirical guidance, election officials across the United States have charted starkly different approaches to election reporting. A number of election officials now affirmatively release individual ballot records---for example, the states of Alaska, Georgia, and Maryland, as well as the counties of San Francisco, California, El Paso County, Colorado, Leon County, Florida, Ada County, Idaho, Tarrant County, Texas, and Dane County, Wisconsin \parencite{ak-cvr,ga-ballotimage,md-cvr, sf-cvr, co-cvr, atkensonLeon, ada-county-nyt, tx-cvr, dane-cvr}---sometimes explicitly with the aim of shoring trust \parencite{ada-county-nyt}.
But many other election officials have prohibited the release of ballot records. 
For example, election officials in Indiana, Missouri, and South Carolina have all denied requests for ballot records \parencite{rivas2022, yorkSC2022, SCAG}, and the North Carolina legislature recently turned back a legislative effort to publish them \parencite{nc-leg-cvr}. 

Election officials and legislators alike have come to different conclusions in part because of an empirical disagreement about the consequences of public disclosure \parencite{fifield2023}. 
For example, jurisdictions which disclose ballot records explain that  ``ballots are anonymous and not connected to voters'' \parencite{Beecher_2023} or ``there [is] nothing linking the data to individual voters'' \parencite{McDonnell_2023}. 
In contrast, jurisdictions which forbid disclosure emphasize that  ``the release of [ballot records] would likely lead to [the] identification of voters'' \parencite{SCAG2} or would ``threaten[] anonymity'' \parencite{hobbs2023AZ}. 

To inform the public debate, we provide the first empirical assessment of the privacy costs of transparent election results, from the aggregate to the individual level. We focus on what we term \emph{vote revelation}. In short, as explained more below, we define vote revelation as linking vote choices to voters. Although a limited set of academic studies and reports have already pointed to the potential for vote revelation, those studies do not go on to measure revelation in an actual election \parencite{adler2013ballots, Colorado2018,bernhard2017public, crs}, or, if they do, only measure a special case of it \parencite{clark2021}.

We use Maricopa County's 2020 general election as our primary case study. Our main finding is that election results could reveal at least one vote choice for approximately 0.00009\% of voters if results were aggregated by precinct,  0.05\% if results were aggregated by precinct and method (the county's current practice \parencite{maricopa-canvass}), and 0.17\% if results were reported by releasing individual ballots. 

\subsection{Framework for Vote Revelation}

To understand the debate about ballot records, it is helpful to start with two background facts about election administration. First, ballots are anonymous, i.e., no names are attached to the actual ballot. Second, the so-called ``voter file'' is not. Instead, it lists all persons by name who are registered to vote in a jurisdiction. It typically includes both where registrants live and if they have voted. In almost every state, voter records in the voter file are public \parencite{eac-voterfile}. In other words, who citizens vote for is private, but whether they register and vote is public.  

\begin{figure}[bt]
\caption{\textbf{Schematic of Potential Privacy Violation by Vote Revelation.} A Venn diagram of how quasi-identifiers could link vote choice to personally identifying information. \label{fig:schematic}}
\centering
\includegraphics[width=0.6\textwidth]{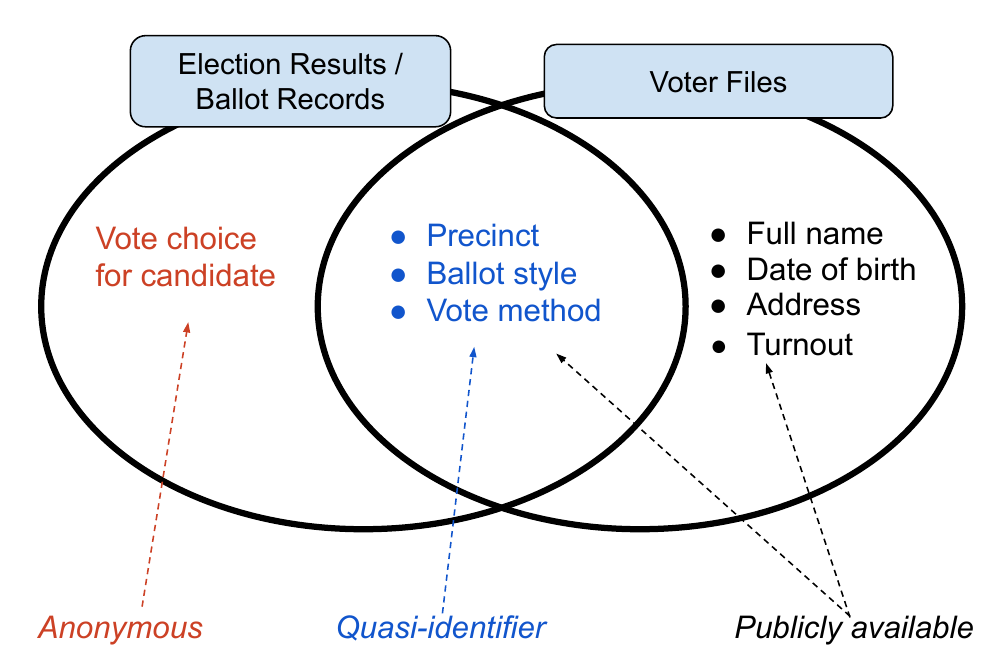}
\end{figure}

One specific concern is that a vote choice on an anonymous ballot will be uniquely linked to the voter’s name in the public voter file---a phenomenon we term ``vote revelation.'' The Venn diagram in Figure \ref{fig:schematic} illustrates how vote revelation is possible. In particular, both ballot records (left) and voter records in the voter file (right) share overlapping information, shown at the intersection of the diagram. The overlapping information includes a voter's precinct, ballot style, and vote method.

Following \textcite{ldiversity2007}, we refer to these overlapping variables as quasi-identifiers. 
They are \emph{quasi}-identifiers, rather than personal identifiers, in the sense that they are not necessarily unique identifiers.
Instead, they narrow down which voter might have cast which vote. 
This setup thus mirrors common studies in the privacy literature, where scholars have shown how an analyst might triangulate multiple data sets with overlapping information to gain information not intended to be shared \parencite{ohm2009broken,sweeney2017re}. While the privacy literature would generally describe this as re-identification \parencite{wood2018differential}, we prefer the term revelation to emphasize the fact that ballots are anonymous.

The quasi-identifiers of precinct, ballot style, and vote method play an important role in both election reporting and election administration. The quasi-identifiers are available in ballot records because they define 
the reporting units for aggregating election results. 
For example, in order to report election results by precinct, election officials need to know the precinct in which each ballot was cast. 
More generally, any reporting unit for election results is defined by some combination of the quasi-identifiers on ballots, either precinct, ballot style, or vote method. 
The same quasi-identifiers are also available in voter files because they define where voters are assigned to vote (a precinct), which set of contests they are eligible to vote in (a ballot style), and how they participate (a vote method).  
Voters in the same precinct may be eligible to vote for different contests and thus have different ballot styles (Figure \ref{fig:balstyle_def}).

\subsection{Types of Revelation}

\begin{figure}[hbtp]
\caption{\textbf{How Unanimous Election Results Reveal Votes.} In this example, 30 voters vote in two contests (President and a tax referendum) and the reporting units for results are defined by precinct $\times$ vote method $\times$ ballot style. \label{fig:schem-unanim}}
\centering
\includegraphics[width=\linewidth]{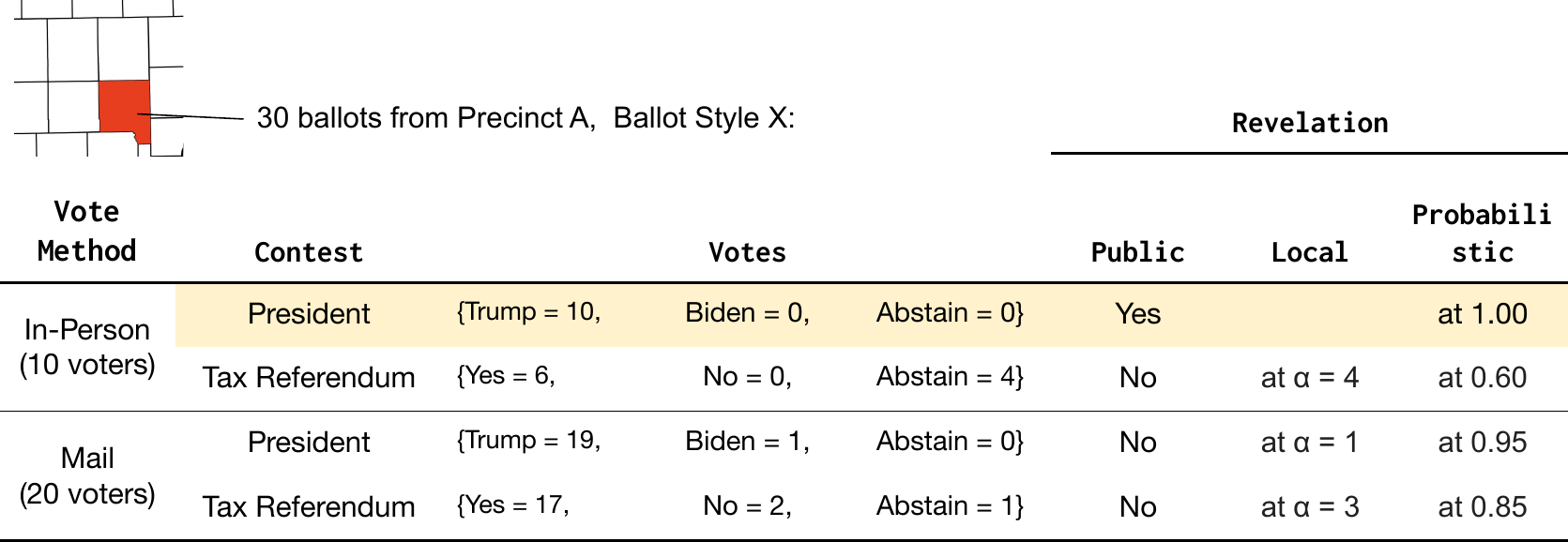}
\end{figure}

Building on the privacy literature discussed in Materials and Methods, we distinguish between three overlapping types of vote revelation: (1) public revelation, (2) local revelation, and (3) probabilistic revelation. Below, we use Figure \ref{fig:schem-unanim} to illustrate each type of revelation using a hypothetical election. 

\paragraph{Public Revelation}

A vote can be \emph{publicly revealed} when all voters with the same quasi-identifiers are unanimous in their vote choice. In the privacy literature, the mechanism for public revelation is thus known as a homogeneity attack \parencite{ldiversity2007}.

In  Figure \ref{fig:schem-unanim} above, the presidential choice of in-person voters is vulnerable to public revelation because all in-person voters were unanimous. An analyst could use the public voter file to find the names of all in-person voters who are registered in precinct A, assigned to ballot style X, and recorded as voting in-person, and know with certainty how they voted for president. 

Because revelation is driven by unanimity, revelation is more likely, all else equal, when election results are reported in reporting units with fewer voters. 
Formally, Equation \ref{eq:model0} shows that the expected number of publicly revealed voters in a reporting unit, which we denote as $R$, is a function of the size of the reporting unit, $N \geq 1$, and the lopsidedness of the contest,
\begin{align}
\mathrm{E}[R] = N\sum^H_{h=1}w_h^{N}\label{eq:model0}
\end{align}
where $w_h$ is the \emph{a priori} probability that a randomly selected voter in the reporting unit supports candidate $h \in \{1, ..., H\}.$ Thus each $w$ is bounded between 0 and 1 (inclusive) and sums to 1. 
In the special case of singleton reporting units (i.e., $N = 1$), we see that a voter's choice is always revealed (\(\mathrm{E}[R] = \sum^H_{h=1}w_h = 1\)). 
Equation \ref{eq:model0} quickly becomes monotonically decreasing in $N$ after a tipping point.
The expected revelations also decrease as races become less lopsided and more evenly contested (e.g., $w_1 = 0.5, w_2 = 0.5$).
In the typical two-candidate uncompetitive contest in which the probability that a randomly selected voter supports the leading candidate is $w_1$ = 0.7 and there are no abstentions, expected revelations drop below 0.01 after $N = 22$ and below 0.001 after $N = 29.$
Supplemental Material \ref{sec:model-more} generalizes the statistical model to allow for abstentions.

\paragraph{Local Revelation}

Votes that cannot be revealed publicly can still be revealed locally. Local revelation can occur if there is private information available about some individual vote choices in the reporting unit, or what privacy scholars refer to as a ``background knowledge attack'' \parencite{ldiversity2007}. 
For example, in Figure \ref{fig:schem-unanim},
the single Biden voter who cast a mail ballot could deduce with certainty that 19 of her neighbors must have voted for Trump.
However, each Trump vote would \emph{only} be revealed to the one voter who did not vote Trump, not the public as a whole.

The extent of local revelation depends on the number of people sharing their private vote choice.
More formally, we denote the number of such collaborators required for local revelation by $\alpha$. The local revelation is only to those groups of $\alpha$ voters whose candidate preferences differ from those of the remaining voters and is only possible in practice if those groups of $\alpha$ voters are able to credibly share their vote choices with each other. 

Equation \ref{eq:model0}, above, can be modified to consider the probability of a $\alpha$-level revelation. Instead of computing the probability of $N$ out of $N$ voters making the same choice, Supplemental Material \ref{sec:model-more} computes the probability of at least $N - \alpha$ out of $N$ voters making the same vote choice using the binomial coefficient.

\paragraph{Probabilistic Revelation}

It is an open question as to whether near-unanimous revelation should also be considered a form of vote revelation. To continue the example in Figure \ref{fig:schem-unanim}  above, an outside analyst who observes the reporting unit with 19 Trump votes and 1 Biden vote, and no other information, would predict that the probability any voter in the unit could be a Trump voter is 95\%. 

Some definitions of statistical privacy do consider near-perfect probabilities of revelation as a successful privacy attack \parencite{hotz2022balancing,duncan1989risk,kenny2023comment}. 
These approaches consider an update in an analyst's beliefs about the values of the sensitive items (here, vote choices) from their prior belief (e.g., based on the voter's quasi-identifiers or partisan affiliation) as privacy leakage. 
In this study, we consider a vote choice probabilistically revealed if it is assigned a (posterior) probability of over 0.95 under the assumption that every voter in each reporting unit is equally likely to have cast each ballot.
We report this operationalization of probabilistic revelation for context and return to its relevance in the discussion. 

\paragraph{Not Vote Revelation}

In Materials and Methods, we also discuss several related privacy concerns that we do not consider to be vote revelation. The most important caveat is that we do not consider a voter voluntarily revealing their vote choice to be vote revelation. We also do not consider the possibility of vote revelation from other potential quasi-identifiers on a ballot, such as voter language or time-stamps. In general, we set these concerns aside because they are not the focus of election officials who are grappling with how to report election results. Instead, these concerns can be addressed outside election reporting.

\subsection{Implications of Revelation}

Based on our framework, the mechanism for vote revelation has two implications for those concerned with privacy violations in election reporting. 

First, individual ballot records do not necessarily reveal more information about how any particular person voted than aggregate election results. That is because the quasi-identifiers that generate vote revelation are available in both individual ballot records and aggregate election results. As in Figure \ref{fig:schematic}, individual ballot records typically have quasi-identifiers for precinct, ballot style, and method. As a result, from the perspective of vote revelation, releasing individual ballot records is equivalent to releasing aggregate election results by precinct, style, and method. Ballot records do not reveal more information simply because they show how a single voter voted on \emph{all} contests. 
To return to Figure \ref{fig:schem-unanim}, even if we had the individual ballots for the ten in-person voters who voted for the same candidate for President, we could still not identify which of those ten voted Yes on the tax referendum.

Second, the extent of vote revelation will vary as a function of both administrative choices about the size of reporting units as well as political behavior. 
For one, we expect primary elections to generate more potential revelation than general elections. In general, fewer voters participate in primaries but the number of reporting units---such as a precinct---typically stay the same. As a result, primaries typically have fewer voters in each reporting unit, mechanically increasing the likelihood of revelation. Primaries may also be more likely to feature lopsided contests, which further contribute to unanimous reporting units. For another, we expect that elections that include many different contests will create more revelations (all else equal) than elections (such as a special election) involving just one contest because the former requires more distinct ballot styles.
In addition, it is not clear whether rural or urban jurisdictions will feature more revelation. Rural jurisdictions will typically have fewer voters per reporting unit in order to accommodate the geographic dispersion of voters. But urban jurisdictions are more likely to have unanimous election results because of political homophily \parencite{brown2021measurement}. 
Finally, there are a set of idiosyncratic factors that influence revelation but do not neatly map onto types of elections or jurisdictions. For example, how jurisdictions report provisional ballots, which are relatively rare, will affect the number of voters per reporting unit.

In the next section, we present a case study of vote revelation along the continuum of election reporting and then assess the generalizability of our findings.

\section{Results} 

We use Maricopa County, Arizona, as a case study to quantify the extent of vote revelation in the reporting of election results. Following the 2020 presidential election, Maricopa County became the most prominent site in the battle over public access to ballots. Maricopa County is also closely divided by partisanship, home to millions of people, and features a range of geographies from urban, suburban, and rural areas. As we describe in more detail in Materials and Methods, we use a particular type of individual ballot record, known as a cast vote record. These cast vote records include the necessary information on both quasi-identifiers and vote choice to allow us to reconstruct election results at any common level of aggregation. 

We first calculate the magnitude of vote revelation in Maricopa County across the continuum of election reporting. We then decompose vote revelation to consider both how to reduce revelation and the potential harm of revelation. Finally, we assess how similar Maricopa County's 2020 general election is to other elections and other jurisdictions.   

\subsection{Prevalence of Vote Revelation}
\label{sec:results-prevalance}

Table \ref{tab:counts_revealed} presents our main finding. The outcome of interest is whether a voter would have \emph{at least one} of their votes revealed in a contested election. Our choice of outcome serves as the worst-case scenario for the magnitude of vote revelation. The top panel presents public vote revelation, the middle panel considers local revelation, and the bottom panel reports probabilistic revelation, while the different columns correspond to different reporting units.

\begin{table}[tb]
\caption{\textbf{Revelation by Aggregation of Election Results}. The number of voters with a revelation in at least one out of the roughly 60 contests on the ballot, according to various definitions of revelation, by level of reporting unit. 
Parentheses translate the number of voters with at least one revealed vote into a percentage of total voters.
\label{tab:counts_revealed}}
\centering
\includegraphics[width=0.8\linewidth]{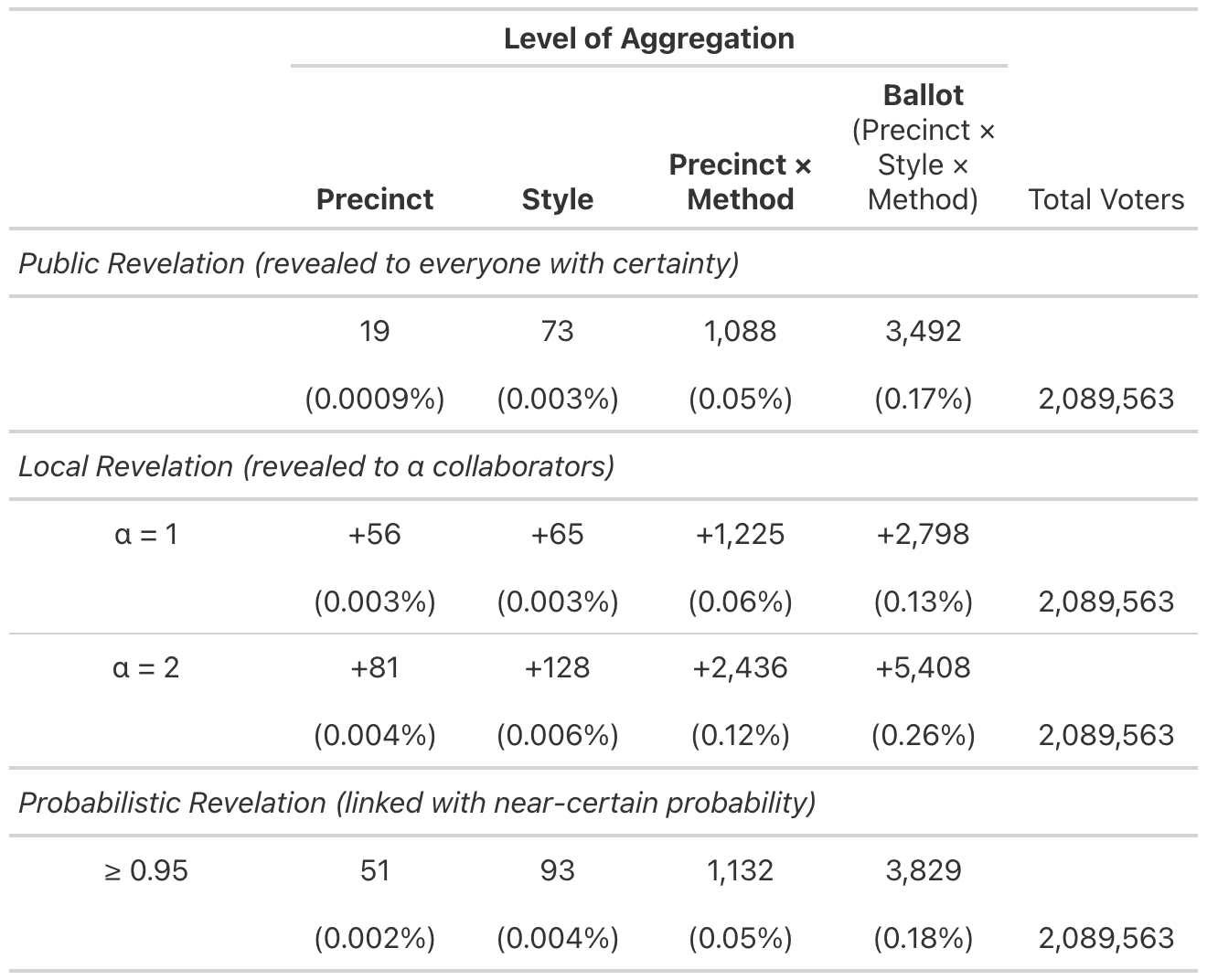}
\end{table}

The first cell in the top panel shows that 19 voters could have at least one of their votes publicly revealed if the 2020 general election results were reported at the precinct level. 
This is 0.0009\% of the more than 2 million voters who participated in Maricopa County's election, or less than one-one hundredth of one-tenth of one percent. 
If the same quantity were reported at the precinct-method level, as in Maricopa County's official canvass \parencite{maricopa-canvass}, 1,088 voters (0.05\%) could have at least one of their vote choices revealed. Finally, in the most granular reporting unit we consider, where election officials release individual ballots, 3,492 voters, or 0.17\% of Maricopa voters, could have at least one vote choice revealed.

The middle panel of Table \ref{tab:counts_revealed} separately reports the increase in local vote revelation relative to public vote revelation. 
The table uses $\alpha$ to define the degree of local revelation.
When $\alpha=1$, every person knows how they themselves voted, but does not share how they voted with anyone else. 
In that case, we find that 56 additional voters would have at least one of their vote choices locally revealed with precinct-level election results. 
The case of $\alpha=2$ might correspond to a household within a precinct, where a couple shares their vote choice with each other. 
We find that 81 more voters (which includes the 56 voters under $\alpha = 1$) will have their vote locally revealed in this case.

Even if we relax the unanimity requirement, the bottom panel shows that the prevalence of revelation increases only modestly. The bottom panel considers probabilistic revelation. Overall, fifty-one voters cast a vote in precincts where the voteshare for at least one contest was at least 95\%.  
Relaxing the threshold from unanimity to 95\% is equivalent to considering local revelation with $\alpha = 1$ in a reporting unit with $N = 20$ voters. 
The increase in probabilistic revelation is more muted at more granular levels of reporting because those reporting units have fewer voters to begin with.

\subsection{Drivers of Vote Revelation}

To consider how to reduce revelation, Table \ref{tab:counts_revealed_method} decomposes the public vote revelation in Table \ref{tab:counts_revealed} by vote method and ballot style. It reveals that the vast bulk of revelation comes from how Maricopa County reports votes on provisional ballots and ballots limited to federal offices.

Federal law generally requires election officials to offer some voters either a provisional ballot or a ballot limited to federal offices \parencite{uocava, hava-provisional}. Provisional ballots may be offered, for example, when an individual may not appear in a precinct's pollbook. Further, federal law also requires that some voters be able to vote for federal offices, even if they are ineligible to vote for state offices. In Arizona, voters must show proof of citizenship to vote in state elections \parencite{az-law}. But the state law does not apply to federal offices \parencite{intertribal}. Instead, federal law permits any voter who attests to being a citizen to be able to vote for federal offices. For this reason, Arizona has a particularly large number of federal-only ballots. 
Maricopa County treats provisional ballots as separate vote methods and federal-only ballots in different precincts as separate ballot styles.

\begin{table}[t]
\caption{\textbf{Revelation by Vote Method and Ballot Style Type}. 
Number of voters with at least one contest subject to public revelation in the ballot-level reporting regime, for each vote method or ballot style type.}
\label{tab:counts_revealed_method}
\centering
\includegraphics[width=0.7\linewidth]{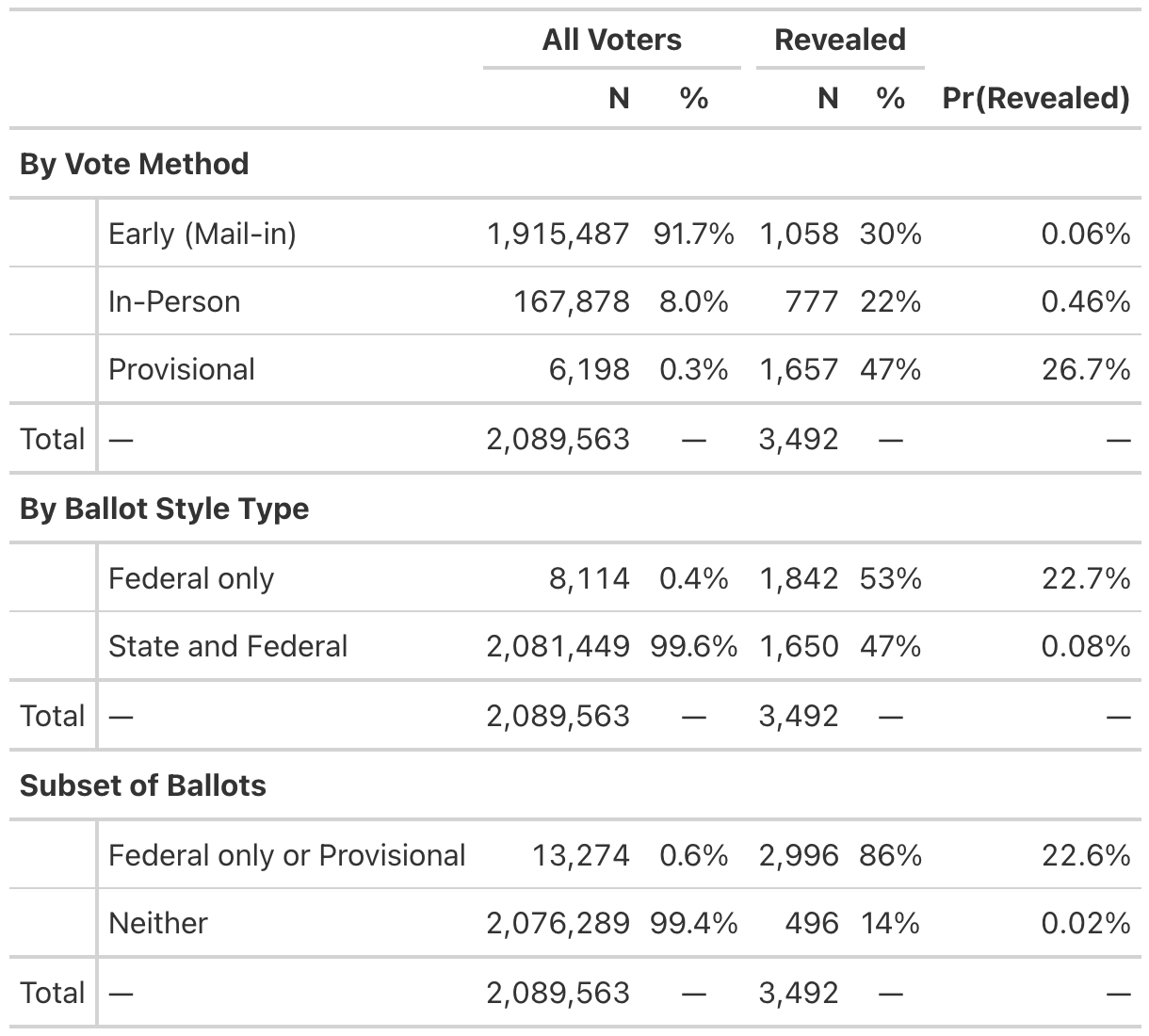}
\end{table}

Table \ref{tab:counts_revealed_method} compares the number of total voters and publicly revealed voters by vote method or ballot style type. The first panel shows that provisional ballots are only 0.3\% of all ballots cast, but they are nearly \emph{half} of the ballots with at least one-contest with a revealed vote. 
Further, the probability of revelation given that a voter casts a provisional ballot is 26 percent. 
The risk of revelation is therefore over 400 times more likely among provisional voters than voters who vote by mail, which is the most common vote method. Similarly, the second panel shows that federal-only ballots are only 0.4\% of ballots cast, but also account for more than \emph{half} of all revealed ballots. The risk of revelation is again highly disproportionate.
Finally, the third panel shows that votes on ballots which are \emph{either} provisional or federal-only are particularly likely to be revealed. Of the 3,492 ballots with at least one vote revealed, 2,996, or 86\%, come from provisional or federal-only ballots.

Provisional ballots and federal-only ballots are the main drivers of vote revelation in part because they are rare. In fact, less than 1 percent of voters end up using either a provisional or a federal-only ballot. As a result, both the provisional ballot voting method and the federal-only ballot style drastically shrink the size of the relevant reporting unit, making unanimity (and thus revelation) more likely. 

\subsection{Patterns of Vote Revelation}
\label{sec:results-patterns}

We also consider whether revelation could disproportionately affect certain types of voters. These patterns may be relevant to public officials who are considering other dimensions of election reporting, beyond the extent of revelation.

\begin{figure}[tbh]
\caption{\textbf{Contests with More Revelation}. A boxplot showing the fraction of public revelations in the ballot-level reporting regime, excluding federal-only ballots. Each point represents a contest. 
The solid bars indicate the median, the box indicates the first and third quartile, and the whiskers extend to 1.5 multiplied by the interquartile range.}
\label{fig:hist_unanim}
\centering
\includegraphics[width=0.75\linewidth]{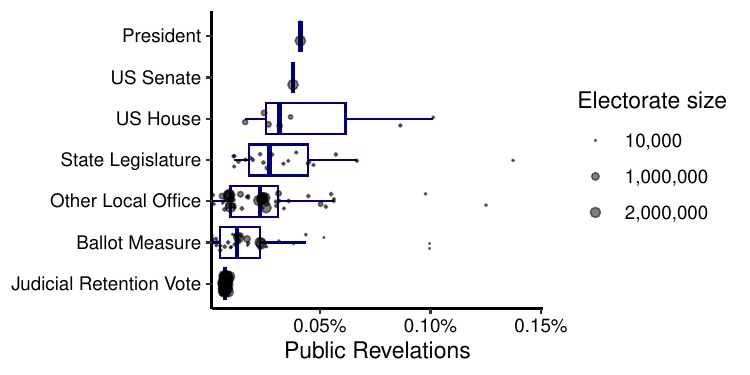}
\end{figure} 

\begin{table}[tbh]
\caption{\textbf{Revelation by Presidential Vote Choice}. 
Number of voters with at least one contest subject to public revelation in the ballot-level reporting regime, for each presidential candidate's supporters.}
\label{tab:counts_revealed_president}
\centering
\includegraphics[width=0.7\linewidth]{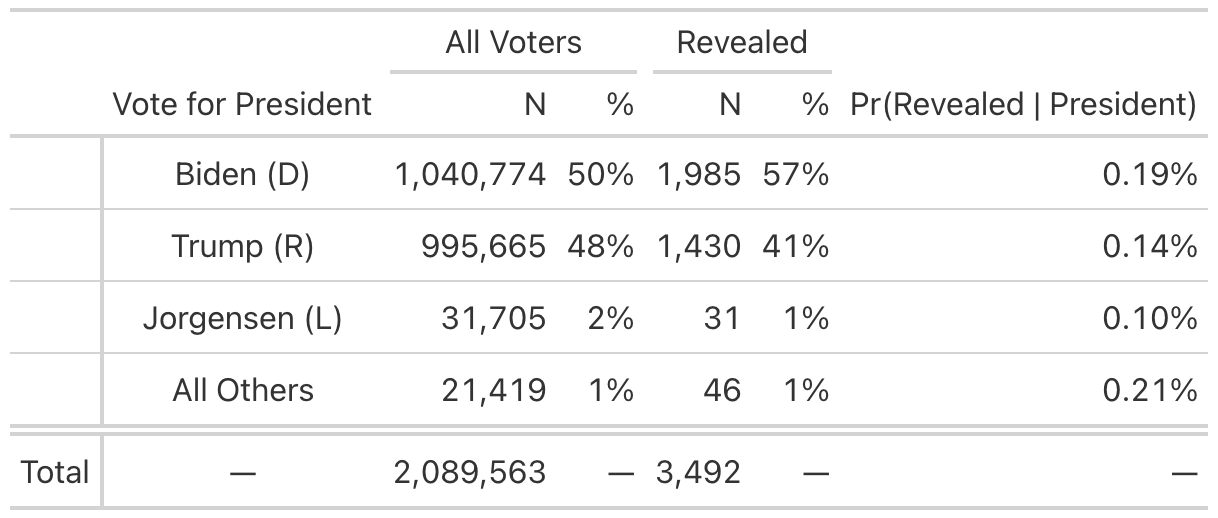}
\end{table}

\paragraph{By Contest} 
Figure \ref{fig:hist_unanim} shows that top-of-the-ticket offices have a higher prevalence of vote revelation than down-ballot offices.
For example, the median judicial retention contest has a potential revelation rate of 0.000012\%, which is three orders of magnitude smaller than the comparable revelation rate for President. 
Revelations track salience because voters tend to abstain in low-salience, down-ballot contests like ballot measures and retention elections.
In fact, Figure \ref{fig:undervotes-lopsided} shows the undervoting rate for a judicial retention election is close to 40 percent, while the undervoting rate for President is only 0.4 percent.  
As a result, even though down-ballot offices are less likely to be competitive, the high abstention rates effectively prevent unanimity. 

\paragraph{By Voter Partisanship} 

A related concern might be that revelation could fall disproportionately on people with certain partisan preferences. 
Table \ref{tab:counts_revealed_president} 
reports the conditional probability that a voter has any vote revealed given their presidential vote choice.
Biden voters are most vulnerable to any-contest revelation (0.19\%), while third-party Libertarian (Jorgensen) voters are least likely (0.10\%). 
There is differential revelation because votes for the third-party candidate for President are not as clustered by geography as votes for the major-party candidates. There are no pockets of neighborhoods that heavily favor Libertarians, whereas some urban precincts may heavily vote for Democratic candidates. 

\paragraph{By Popularity} 

Yet another concern is that revelation could
fall disproportionately on people who hold unpopular views. 
As detailed in the Supplemental Material \ref{sec:radius-appendix}, we compare the presidential vote choice of each revealed voter to the choice of other voters living in the surrounding geographies.
For each fixed radius from the revealed voter, we compute the degree of agreement between revealed voters and all other voters.

\begin{figure}
\caption{\textbf{Revelation of unpopular vote choices}. For each revealed vote choice at the precinct-level or ballot-level, we display what percent of the revealed voter's neighbors who share that vote choice, where neighbors are defined by geographical distance. Solid lines show average agreement by revealed candidate.}
\label{fig:radius}
\centering
\includegraphics[width=0.85\linewidth]{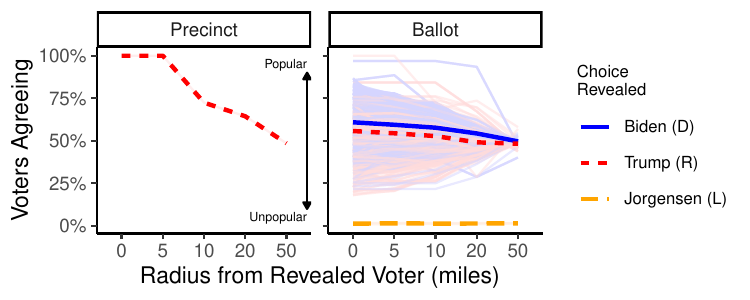}
\end{figure}

Figure \ref{fig:radius} summarizes our agreement measure when results are reported at either the precinct-level (left) or ballot-level (right). An important difference between the two reporting units is that a precinct is a geographic unit while a ballot contains a mix of geographic and non-geographic units, such as vote method.

The left facet has a single line because Maricopa County had only a single precinct whose votes for president were unanimous. 
By definition, the agreement value is 100\% at $x = 0$ miles since revelations arise from unanimous results.
At a $x = 10$ mile radius, the agreement value drops to about 75\%, meaning that 3 out of 4 voters in a 10 mile radius from the revealed voters agree with the revealed voters' choice.

The agreement value on the right panel does not begin at 100\%
because one vote method within a precinct may be unanimous while another method in the same precinct is often not.
The solid lines show the weighted mean agreement with revealed votes for each candidate.
The revealed votes for Biden and Trump tend to reflect the majority, not the minority, view. 
However, the Libertarian voters who are revealed are distinctly unpopular at any radius because Libertarian preferences are not clustered by geography or method. 

\subsection{Case Study in Context} 

We expect the prevalence of vote revelation to vary both across elections and across jurisdictions, in part because of differences in the number of voters per reporting unit. Further, while the patterns of revelation by contest are likely consistent across jurisdictions, the other patterns of revelation by preference or popularity could vary based on the extent of partisan homophily. For these reasons and others, we compare Maricopa County's 2020 general election to other elections and to other jurisdictions.

\paragraph{Different Elections}

Table \ref{tab:revealed_2022pri} compares Maricopa's 2020 general election and its later 2022 primary election, using similar cast vote records. 
The 2022 primary election only had a third of voters compared to the 2020 general election, but it had the same number of precinct reporting units.
Consistent with our statistical model, voters in the 2022 primary election were 2 to 5 times more likely to have their vote choices revealed than in the 2020 general election.

\paragraph{Different Jurisdictions}

Analyzing revelation in the 2020 general election across all jurisdictions nationwide is beyond the scope of this study, particularly for individual ballot records. Instead, we consider the relative difference in the size of reporting units across jurisdictions. Given both our theoretical framework and empirical results, we focus specifically on the number of voters in reporting units with few voters, since these voters are particularly vulnerable to revelation. 

Table \ref{tab:snyder-stats} makes use of a recently published dataset of cast vote records for the 2020 general election from 160 counties \parencite{conevska2024partisan}, as further detailed in Supplemental Material \ref{sec:snyder-collection}.
These counties are scattered around 20 states and are not a representative sample of the U.S. population, in part because many jurisdictions do not release cast vote records. 
They are collectively slightly more non-White and 3 percentage points more Democratic-leaning than the country as a whole (Table \ref{tab:snyder-demographics}).
Nonetheless, the available comparison is informative for contextualizing the prevalence of revelation in Maricopa County.

Overall, Maricopa County had relatively more voters per precinct in the 2020 general election but relatively fewer voters per some ballot styles, particularly the federal-only ballots discussed above. Focusing on the precinct-level first, Maricopa County had only 1 voter per 100,000 in precincts with less than 10 voters and only 3 per 100,000 in precincts with less than 30 voters. In comparison, the \textcite{conevska2024partisan} counties had 12 voters per 100,000 in precincts with less than 10 voters, and 49 in precincts with less 30 voters. The comparison set of counties also allow us to examine number of voters in ballot-equivalent reporting units, defined by precinct, ballot style, and vote method. Maricopa had 381 voters per 100,000 in ballot-equivalent units of less than 10 voters, while the  \textcite{conevska2024partisan} counties had just 158 voters per 100,000. 

Given these relative differences in the size reporting units, we expect the precinct-level revelation we find in Maricopa County may be closer to a lower bound of revelation across the country, while the ballot-level revelation may be closer to an upper bound. However, analyzing revelation in the \textcite{conevska2024partisan} data is beyond the scope of this study, as they did not validate their data against the certified election results, as we have carefully done for Maricopa County. While validation was not necessary for their study of ticket splitting, validation is particularly important for vote revelation because public revelation is driven by the unanimity of results. Supplemental Material \ref{sec:medsl-construction} instead compares revelation across 47 states in a validated dataset of 2020 presidential election results at the precinct level \parencite{baltz2022precinct}. 

\section{Discussion}

Given our results, we finally address the future of election reporting. 
We first review the evidence on the value of transparent election results, drawing from the political science literature.
We then evaluate current ex-post approaches to reducing revelation and propose additional ex-ante approaches. We finally consider the limitations of any empirical case study. 

\subsection{Benefits of Transparency} \label{sec:transp-benefits}

A growing literature in political science finds support for theories that more transparency in election reporting can reduce voter fraud or increase voter trust. 

For one, granular election results might reduce fraud because they facilitate the detection of voting irregularities. In fact, the growing body of so-called election forensics uses granular election results precisely for that reason. For example, in a classic study, \textcite{wand2001butterfly} used multiple sources of granular election results, including cast vote records, to show that the effect of poor ballot design was large enough to swing the winner of the 2000 Presidential election \parencite[see also][]{morseworking}. More recent studies similarly rely on results reported by vote method and precinct to dispel claims of election fraud \parencite{goel2020one,eggers2021no,Bafumi2012,cantu2019fingerprints} or detect actual election fraud  \parencite{herron2019mail}. 

Consistent with election forensics, international organizations monitoring elections recommend that countries publish more granular election results to reduce fraud. 
The findings of various political science studies support that recommendation. 
For example, a field experiment in the 2010 Afghanistan parliamentary elections found that notifying polling station managers that their initial vote tallies will be photographed and made public substantially reduced the manipulation of vote counts  \parencite{callen2015institutional}. 
Further, an over-time study of 125 low- and middle-income countries finds that more granular election results are associated with fewer perceived irregularities \parencite{rueda2023more}.
The 2024 presidential election in Venezuela offers a particularly stark example of the need for transparency. 
In that election, the politically-controlled election board declared that the incumbent had won with 52 percent of the total vote, while witholding polling station-level statistics that showed he had in fact lost the election \parencite{kronick2024}.

More granular election results may also improve public trust by changing perceptions.  
\textcite{rueda2023more} find that more granular results improve expert perceptions of the overall quality of international elections. 
Further, in the contemporary U.S., \textcite{jaffe2023} shows that publishing post-election audits increases the mass public's confidence in both the accuracy and result of the election. 
The effect of publicly releasing audit results is large--- \textcite{jaffe2023} finds that its effect on the public's confidence in the audit is about as large as the effect of the declared winner in the audit being a co-partisan. 
It is possible the effect might be even larger if the \emph{ballots} being audited were public, too. Then everyone could participate in the process, skeptic or not. 

Granular election results have also become central to a range of other aspects of election research.  
For example, granular election results are relied upon to demonstrate racial polarization relevant to the enforcement of current voting rights protections.\parencite{greiner2007ecological}. 
Cast vote records in particular also allow better measurement of ticket splitting and voter behavior in ranked choice elections \parencite{kuriwaki2023ticket,alvarez2018low,pettigrew2023ballot}.

\subsection{Ex-post Approaches to Reducing Revelation}

Transparency, of course, has costs as well as benefits.
Election officials are broadly pursing two types of policies to reduce or eliminate vote revelation.
The first approach uses redaction to remove certain information from election results, while the second injects noise into those results. 
These approaches can either target the quasi-identifiers associated with individual ballots or the vote choice themselves.
In general, modifying vote choices has more privacy guarantees than modifying quasi-identifiers \parencite{cohen2022attacks} but runs counter to a fundamental goal of elections to identify a winner. 
\paragraph{Redaction} Several states redact information from all reporting units that have fewer than a certain number of voters. 
For example, by statute, election officials in Nevada, New Mexico, and Florida are required to redact vote method from precinct-method aggregate totals if the number of votes cast in any precinct-method reporting unit is below a defined threshold of either five, ten, or thirty total votes \parencite{nm-law, nv-law, fl-law}. 
In that case, election officials would report results for the particular precinct as a whole rather than separating it by method. 
A similar approach is taken in other democracies too. For example, in Germany, election officials combine any precinct with less than 50 total votes with another precinct. The approach effectively redacts a precinct identifier by reporting two precincts as if they were one \parencite{germany-law}.
Such redaction and aggregation can be an intuitive approach for election officials to reduce revelation from election reporting.

\begin{figure}[t]
\caption{\textbf{Tradeoffs from redaction}. Number of affected ballots from a hypothetical policy of redacting ballots from reporting units with $k$ or fewer voters. $k = 0$ indicates the status quo of no redaction.}
\centering
\label{fig:suppression}
\includegraphics[width=0.75\linewidth]{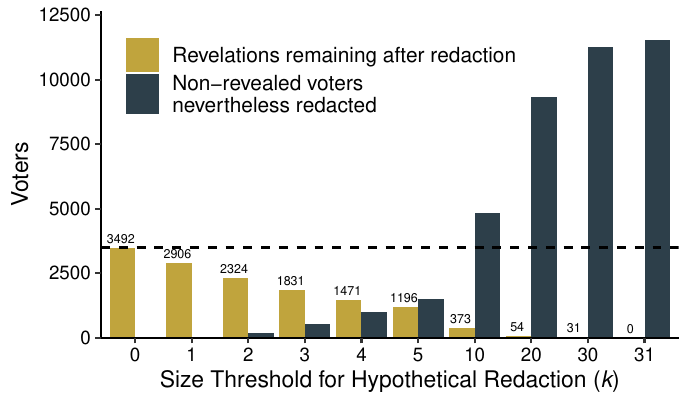}
\end{figure}

Nonetheless, redaction almost always reduces transparency too. Figure \ref{fig:suppression} visualizes the tradeoff by returning to Maricopa County's 2020 general election. The horizontal, dashed line reflects that there are 3,492 votes vulnerable to revelation, using the ballot-level, any-contest specification as in the previous figures. The x-axis varies the threshold for redaction, which we denote by $k$. The y-axis counts the number of voters whose ballots would be subject to some form of redaction if they were part of reporting units with less than $k$ voters. For any particular threshold, the figure distinguishes between voters whose vote is (blue) or is not (yellow) vulnerable to public revelation before redaction. 
 
By definition, a policy of no redaction ($k = 0$) would expose all votes vulnerable to revelation, while a policy of redacting and aggregating ballots in reporting units with just a single voter ($k = 1$) would reduce revelation without affecting ballots not vulnerable to revelation. In order to more substantially reduce revelation, an election official could increase $k$. 
For example, using a threshold of $k = 5$ or fewer voters could reduce revelation by 65\%, but not eliminate it.
However, the number of voters whose ballot is redacted despite not being vulnerable to revelation would increase as well.  
In our example, Maricopa County would need to set $k$ to 31 to eliminate revelation. As a consequence, though, more than 11,000 voters whose ballots are not vulnerable to revelation would also be subject to some form of redaction.

More generally, the harm of redaction depends on both the information redacted and the other information about election results available in the jurisdiction. For example, many Colorado counties redact the vote choice from ballot records of voters in reporting units of less than 10 voters before they publicly release a cast vote record database \parencite{kuriwaki2024cast, co-redaction}.
In practice, New Mexico and Nevada appear to redact the number of votes that each candidate receives in small reporting units, rather than the vote method. 
Redacting the vote choice means the public cannot verify which candidate got the most votes, at least in that specific set of election results. 
However, all of these states also report unredacted vote counts at higher levels of geography (e.g. counties and statewide). 

\paragraph{Extending Redaction}

While the policies above deploy redaction based on the size of reporting units, a similar approach could instead target the particular types of ballots which are disproportionately likely to generate revelation. Recall that 86 percent of the potential revelations from releasing individual ballots in Maricopa County's 2020 general election came from provisional and federal-only ballots (Table \ref{tab:counts_revealed_method}). 

The two types of ballots call for somewhat different treatments but simple reporting adjustments could prevent most revelation. 
For one, election officials do not need to identify provisional ballots in election results. While there is a clear benefit to identifying voters whose provisional votes need to be cured, once those ballots are counted, the benefit to identifying which ballots were provisional ballots is less clear. 
Election officials could instead modify vote method for provisional ballots to simply identify that it was cast in person, as provisional ballots typically are.
For another, election officials could report provisional ballots (or federal-only) ballots in a separate non-geographic precinct.
The upshot would be larger reporting units and thus less revelation.

\paragraph{Noising}
\label{sec:discussion-noising}

Beyond redaction, North Carolina has charted a unique approach of adding noise to election results. 
In short, the state adds or subtracts an undisclosed random number to some reported election results. 
The policy --- what we term \emph{noising} --- is consistent with the calls of many contemporary privacy scholars in the context of administrative data generally. 
But the policy is a poor fit for election administration.
 
In theory, adding random noise can be a powerful privacy protection mechanism because it makes sensitive data more resistant to background knowledge attacks, while largely preserving the ability to draw correct inferences about the structure of the underlying data. Such privacy protection satisfies what scientists refer to as the differential privacy criterion \parencite{dwork2006differential,evans2019statistically,abowd2023,kenny2023evaluating}. However, valid inferences are only possible when the analyst has full knowledge of the noise injection mechanism and North Carolina has yet to publicly document its random noise algorithm.  

More generally, differential privacy faces conceptual challenges when it comes to election reporting.
Classic differential privacy requires that the reported result of every query to a confidential database be effectively equally likely regardless of whether any individual is removed from the database.  
The most important query for election results is simple: which candidate won the most votes? 
The answer to this query must be reported with perfect fidelity in close elections.
But as we explain in Materials and Methods, differential privacy requires that the answer not substantially depend on any one vote, or, put another way, that we pick election winners in a way that no single vote could ever matter. 
In North Carolina, the state board warns that a candidate who appears to have won the most votes in the noised results may not be the one who was the actual certified winner \parencite{nc-readme}.

Many of these challenges might be solved by advances in privacy algorithms. 
Some recent variants of differential privacy do incorporate constraints so that, in our case, the certified election totals at the constituency level would not be altered \parencite{gao2022subspace, subspaceDP2023}. 
However, these methods are complex, require expertise and documentation, and are yet to be implemented in election administration. 
 
\subsection{The Importance of Ex-Ante Electoral Design}

Ex-post approaches to eliminate vote revelations after an election are inherently second-best.
Instead, election officials should also consider \emph{ex-ante} approaches to reduce revelation before conducting the election.

Ex-post approaches such as redaction, coarsening, and noising only protect voters’ from having their vote choices revealed to the public. They do not protect from voters from having their choices revealed to election officials themselves. The secret ballot, though, was developed in part to protect the privacy of voters' political preferences from the government \parencite{keyssar2009right,mares2015open}. %
For example, if vote choices in a given reporting unit turn out to be unanimous, that information could be revealed to election officials themselves before they perform any redaction or noising. In other words, if the reporting of election results violates the secret ballot, it could do so regardless of whether the results are made public or not. Further, the ex-post redaction and coarsening of quasi-identifiers is not only non-trivial, but can easily generate the sort of distrust that officials are seeking to reduce.

A more robust secret ballot requires designing election reporting systems that are less likely to produce vote revelations. One way for jurisdictions to reduce the likelihood of revelation is to re-draw political districts so that there are more voters per reporting unit and fewer or no rare ballot styles.
However, some district boundaries may intersect with many other district boundaries and yet not be readily manipulable.
Beyond redistricting, revelations can also be avoided by ``reprecincting'' to reduce the number of precincts (and thus expand the number of voters per precinct). For example, jurisdictions that employ vote centers or vote-by-mail rather than traditional in-person polling stations could define their precincts by ballot style and thus avoid revelations created when precincts are split by ballot style.

A jurisdiction might also modify the information available in the public voter file.
Voters' names and addresses are easily linked to unanimous election results in the U.S. because names and addresses are publicly available in voter files (Figure \ref{fig:schematic}), along with information on whether registrants turned out to vote. 
Limiting that information on the voter file would make vote revelation much more difficult. 
But we expect state legislators are unlikely to adopt the approach, since parties depend on voters' public contact information and vote history to mobilize and target support \parencite{hersh2015hacking}.
Further, it would not address the prospect of revelation to election officials.
A similar but less drastic approach would limit the snapshots of the voter file made available, in effect using the churn of the electorate for privacy protection. 
In conjunction with these approaches, states might also consider centralizing standardizing, and regulating the election reporting process. In fact, some states, such as Washington and Georgia, have already moved in that direction \parencite{green2024foia}.

\subsection{Remaining Concerns}

Election officials faced with navigating the privacy-transparency tradeoff in election results may have other concerns beyond those addressed in our case study.

For example, election officials might worry about the possibility of revelation even if there is not certain revelation. In general, though, probabilistic revelation seems to us to over-state the privacy risks of election reporting. A probabilistic approach fits uneasily in the U.S. election setting, where campaigns are already flush with individually-identifiable information about voters \parencite{hersh2015hacking,rentsch2019elusive}. 
The public voter file typically lists voters' party of registration, and campaign contributions are typically public too. 
Further, there is no clear threshold to distinguish probabilistic revelation. 
Regardless, our results show that the extent of probabilistic revelation at 95\% is similar to the amount of certain revelation (Table \ref{tab:counts_revealed}). 

Election officials may also be concerned about the public's \textit{perception} of privacy violations.
Regardless of the extent of actual revelation, perceptions of revelation may discourage participation \parencite{gerber2013perceptions}. In fact, a substantial number of Americans think their vote choice is not actually secret \parencite{gerber2013there,atkeson2023voter}. However, the vast majority of Americans also report freely sharing their vote choice with others \parencite{gerber2013there}. Nonetheless, it would be concerning if publication of ballot records increases perceptions of privacy risk. Ultimately, future work should consider how best to inform voters about the actual prevalence of revelation.

Separately, election officials might worry that our case study will not generalize to their jurisdiction or to various types of elections. Ultimately, we expect vote revelation will vary by election and jurisdiction. In fact, we showed that revelation is 2-5 times greater in Maricopa County's 2022 primary election than in its 2020 general election. However, we also found that Maricopa County likely had \emph{more} revelation from releasing ballot records than we would expect in other jurisdictions, primarily because of its particular policies for reporting provisional and federal-only ballots, both of which can be easily remedied.

Finally, while election reporting involves balancing privacy and transparency, our study does not quantify the value of transparency. If transparency is of little value, the case for more granular election reporting is necessarily weaker. On the other hand, if transparency is of some value, then taking transparency seriously requires reducing potential revelation from election results. 

\subsection{The Future of Election Reporting}

In this study, we extensively considered the ways in which a voter's vote choice can be revealed through the reporting of election results and measured the actual prevalence of what we call public, local, and probabilistic revelation. 
Conceptually, we have shown that vote revelations are possible in \emph{any} election reporting system that promotes transparency. Empirically, we found that the release of individual ballot records can publicly reveal a vote choice of about two-tenths of one percent (0.17\%) of the voters in a large, diverse county. In comparison, the release of aggregate results by precinct and vote method would reveal a vote choice for about one-twentieth of one percent (0.05\%) of voters.
Further, while Republican and Democratic voters were equally likely to have their vote choices revealed, voters who cast provisional ballots and ballots limited to federal offices were disproportionately more likely to be affected. 

In the coming years, we expect election officials, legislators, and the public alike to continue to debate how to best balance transparency and privacy in election reporting. Choosing one particular reporting regime over another depends on both the empirical realities we detail here as well as normative preferences about the value democracies should place on privacy versus transparency. In that sense, our case study is not designed to dictate any particular policy as much as it offers a framework to arrive at one.

\section{Materials and Methods}
\label{sec:materials-and-methods}

\subsection{Materials}

We use Maricopa County, Arizona, as a case study to quantify the extent of vote revelation in the reporting of election results.

\subsubsection{Case Selection}

We used cast vote records from Maricopa County's 2020 general election and 2022 primary election. For each contest on each ballot, Maricopa's cast vote records provide vote choice, including undervotes or overvotes, vote method (in-person, early, provisional), precinct, and ballot style. Maricopa's cast vote records thus include the necessary information on both quasi-identifiers and vote choice to allow us to reconstruct election results at any common level of aggregation.
We inspected the dataset extensively to verify that it reproduced the result from the official canvass of results. 

In the main text, we analyze contested single-member contests with non-transferable votes. 
We define a contested contest as a contest having at least two candidates listed on the ballot, not counting write-ins, or having at least two options on a referendum. Based on this definition, over 99.5 percent of the ballots cast in the 2020 general election included at least 64 contested contests. 
Figure \ref{fig:undervotes-lopsided} shows a summary of the contests we examine.

\subsubsection{Reporting Units}
\begin{table}[tbp]
\caption{\textbf{Summary Statistics}. Counts and coverage of the Maricopa County data we focus on in this paper. Each row indicates a reporting unit. The column ``unique units'' indicates the number of unique units that exist. Cells show the proportion of voters in small reporting units, which correspond to the empirical cumulative density function weighted by the number of voters in each unit. 
\label{tab:data_maricopa}}
\centering
{\footnotesize

\begin{tabular}{rcccccc}
\toprule
 & & \multicolumn{5}{c}{Voters per 100,000 in units of size:}\\\cmidrule(lr){3-7}
Reporting Unit & Unique units & 1 & $\leq 2$ & $\leq 5$ & $\leq 10$ & $\leq 30$ \\\midrule
Precinct & %
743 & 0 & 0 & 0 & 1 & 3
\\
Ballot Style & %
381 & 0 & 0 & 2 & 5 & 13
\\
Precinct $\times$ Ballot Style & %
1,741 & 1 & 4 & 34 & 124 & 358
 \\ \addlinespace
Ballot equivalent & %
4,397 & 28 & 65 & 182 & 381 & 705
\\
\multicolumn{1}{r}{\hspace{1em} ... \emph{only mail method}} & %
1,722 & 2 & 8 & 47 & 143 & 319
\\
\multicolumn{1}{r}{\hspace{1em} ... \emph{only in-person method}} & %
1,369 & 126 & 256 & 469 & 657 & 1,530
\\
\multicolumn{1}{r}{\hspace{1em} ... \emph{only provisional method}} & %
1,306 & 5,357 & 12,649 & 34,124 & 66,731 & 97,661
\\
\bottomrule
\end{tabular}
}
\end{table}

Table \ref{tab:data_maricopa} summarizes our collection of cast vote records for the November 2020 general election. 
It reports both the number of unique reporting units and the proportion of voters in reporting units with few voters, which corresponds to the empirical cumulative density function weighted by the number of voters in each unit. 
For example, the top row shows that Maricopa County used 743 unique precincts for the election. 
All in all, only 0.003\% of voters (3 out of 100,000) voted in precincts with 30 or fewer voters. 
The most granular reporting unit is precinct $\times$ ballot style $\times$ vote method, which we have established is equivalent for purposes of vote revelation to releasing individual ballots. 
There are 4,397 such reporting units, with 0.7\% of voters  in units with 30 or fewer voters and 0.028\% in singleton units.  
The three rows with ``...'' further subset style-method reporting units by each of the three vote methods.  
Reporting units for provisional ballots have by far the fewest number of voters. Nearly 98\% of voters who cast a provisional ballot are in ballot-level reporting units of 30 or fewer voters (last row; 97,661 out of 100,000 voters).

\subsubsection{Cast Vote Records versus Ballot Images}

Cast vote records are one of two different types of individual ballot records, each with subtly different implications for vote revelation. However, the two records have often been confused in the public debate. 
 
A \emph{cast vote record} is a machine-readable enumeration of a voter's choice across some or all contests on their ballot \parencite{wack2019cast,leingang2022}. Cast vote records are generally stored in familiar data formats such as XML, JSON, or spreadsheets.  
A \emph{ballot image} is a digitally recorded scanned image of each ballot or ballot page. 
Vote scanners can produce both types of individual records. 

From a privacy perspective, the relevant difference is that cast vote records only contain a voter's contest choices, while a ballot image would capture all marks or notations made on the ballot, including those that do not indicate a choice for a particular candidate. 
For this reason, voters are often prohibited from adding identifying marks to their ballot to identify themselves \parencite[see, e.g.,][]{mn-law}, and election officials attempt to redact ballot images with identifying marks \parencite[][see also Figure \ref{fig:leon} for a redacted mark]{Colorado2018}.

Nonetheless, ballot images are more vulnerable to vote revelation than cast records because deterrence from criminal sanctions and redaction by election officials may be ineffective or incomplete. Further, our interest is not in self-revelation, but rather the revelation of vote choices that were not intended to be revealed to anyone. For this reason, we study what the choices themselves can reveal. Those choices are contained in the cast vote records.

\subsubsection{Voter File Data}

The possibility of vote revelation, from either individual cast vote records or aggregate election results, depends on the completeness of the voter file. 
In practice, the public voter file is incomplete.
For one, there is well-known churn in publicly available voter files, which would incidentally prevent revelation \parencite{nyhan2017differential, kim2022voter}. 
Further, Maricopa's Address Confidentiality program removes the records of about 4,000 voters from their public voter file. 
These voters are typically victims of domestic abuse and stalking who are allowed to shield their addresses from public records \parencite{maricopa2022}.
Finally, Maricopa's voter file lumps together in-person votes and provisional votes under the same vote method. 

Nonetheless, we assume that such a complete dataset is available, and measure the amount of \emph{potential} revelation that would be possible with a complete list of voters and their quasi-identifiers. 
The main reason is that election officials, such as those in Maricopa County, retain a list they refer to as the ``voted file'' that is not affected by voter churn, that does not redact confidential addresses, and does not coarsen vote method. 
Anyone can obtain a voted file through an open records request. 
And even if voted files were not available to the public, the state has access to such a file.
Supplemental Material \ref{sec:voterfile} provides more detail on our comparison between this voted file and a commercial voter file.

\subsection{Methods}

Our methodology builds on work in the formal privacy literature and extends it to the unique context of election reporting. 

\subsubsection{Vote Revelation in a Privacy Framework}

The building blocks of an election database include a table of individual ballot records and a table of individual voter registration records.
As Figure \ref{fig:schematic} illustrates, the ballot records indicate each choice made in each contest by each voter in any election, while the voter registration records enumerate which registrants cast a ballot in the same election.
Importantly, the ballot records are anonymous, i.e., no names are attached to the actual ballot, but contain quasi-identifiers (e.g., precinct, ballot style, vote method). 

A formal statement of what we call public revelation is based on a measure called $\ell$-diversity \parencite{ldiversity2007}. A dataset is considered $\ell$-diverse if there are at least $\ell$ distinct ``well-represented'' values of a sensitive attribute (vote choice) in a block of data with a specific quasi-identifier (precinct, style, or method). The concept of $\ell$-diversity contrasts with the older notion of $k$-anonymity \parencite{sweeney2002k}, which holds that a dataset is $k$-anonymous if there are at least $k-1$ individuals with the same quasi-identifier. Even if a dataset is $k$-anonymous, a sensitive attribute can be revealed if all $k-1$ voters with a specific quasi-identifier vote unanimously (in other words, $\ell = 1$ with only one value). However, datasets that are $\ell$-diverse are still not necessarily immune from vote revelation \parencite{cohen2022attacks}. 
Instead, a person who voted themselves can use their own private knowledge to back out how others voted. For this reason, the mechanism for what we call local vote revelation is known as a \emph{background knowledge attack}  \parencite{ldiversity2007}.

\subsubsection{Not Vote Revelation}

We set aside various related privacy concerns in our definition of vote revelation.

\paragraph{Other Quasi-identifiers} First, we do not consider the possibility of vote revelation from quasi-identifiers other than precinct, ballot style, and vote method. 
For example, election officials have voiced concern that including time-stamps on ballots records indicating exactly when each ballot was cast (or in what order the ballots were cast) could be used to reveal votes.
In fact, at least one voting system is known to have inadvertently disclosed the order in which in-person ballots were cast \parencite{dvsorder}.
Relatedly, some states allow for so-called ``retrievable'' ballots. For example, North Carolina marks ballots cast during early voting with an identifier such that it is ``retrievable'' \parencite{nc-retrievable}.

We set aside such concerns arising from other quasi-identifiers because the solution to these concerns is about election administration, not the reporting of election results. To continue the examples above, election officials need not provide a time-stamp of a voter's interaction with a ballot in a publicly available cast vote record. Further, the ballot identifier is not a public record and can only be linked to a voter if a vote is later found to be fraudulent. Otherwise, it is a criminal offense for anyone ``who has access to an official voted ballot or record and knowingly discloses ... how an individual has voted that ballot'' \parencite{nc-crimlaw}.

In contrast, if election officials are required to report precinct-level results, they must \emph{necessarily} associate the voter's precinct with each ballot in order to administer the election. Similarly, information about the ballot style is inherent to each ballot and cannot typically be redacted.
(However, the cast vote records produced by some voting systems do not connect vote choices made by voters across a multiple-page ballot \parencite[see, e.g.,][]{morse2021future}. In this case, ballot style may not be entirely revealed by the cast vote records and, thus, ballot style information is not inherent to cast vote records in those cases.) 
Election officials have also voiced concern about ballots in different languages, but we are not presently aware of ballot language being recorded in a voter file.

\paragraph{Authentication} Our definition of vote revelation also does not consider the possibility of voters using cast vote records to \emph{voluntarily} reveal their vote. 
Prior work has warned that releasing ballot images could allow voters to authenticate their votes by adding identifiable stray marks, which could be captured in ballot images \parencite{adler2013ballots}. 
That concern is also reflected in state law. 
For example, Minnesota requires that a ballot not be intentionally marked with ``distinguishing characteristics'' that render it identifiable \textcite{mn-law}. 
More recent work focuses on how releasing cast vote records in ranked-choice elections could similarly allow each voter to use their complete ranking ordering of the candidates as a verifiable public signal that would nearly always be unique to them \parencite{williams2023votes, ryan2021masked}. 

Nonetheless, we see vote revelation as a more pressing concern than vote authentication in considering the public disclosure of individual ballots. 
For one, vote-buying and vote-selling already violates federal law \parencite{hasen2000vote}. 
Further, vote authentication is often permissible as protected First Amendment speech, at least in some jurisdictions \parencite[see, e.g.][]{rideout}. 
For example, many states permit voters to share a picture of their marked ballot, known as a ``ballot-selfie'' \parencite{ncslselfies}.

\paragraph{Negative Revelation}

Our definition of vote revelation does not include so-called negative revelation: the ruling out of choices that a voter may have made. Negative revelation is often found in granular election results because zero votes might be cast for particular minor-party candidates in a reporting unit. 
Cast vote records can allow more negative revelation than aggregate election results. 
To use the set up of \ref{fig:schem-unanim}, suppose a vote buyer asks one voter to vote for Trump and vote Yes on a referendum, and another to vote for Biden and vote No on the same referendum. However, if the first voter voted [Trump; No], and the second voted [Biden; Yes], the cast vote records would reveal that both voters had reneged, while aggregate results would not. 

\paragraph{Vote Verification} 

Finally, our definition of vote revelation pertains to revelation to others, not self-verification. Cryptographers have designed encryption systems that, when properly implemented, allow voters to verify to themselves that their vote was properly tallied as cast \parencite{bernhard2017public}.

\subsubsection{Prevalence and Patterns of Vote Revelation}

Our main results calculates whether a voter had \emph{at least one} of their votes revealed in a given election. In other words, if we define $r_{ij} = 1$ when voter $i$'s vote for contest $ j \in \{1, ..., K_i\}$ is revealed, and $r_{ij} = 0$ otherwise, we count the number of voters for whom $r_{ij} = 1$ in at least one contest $j$. 
To formalize our quantities of interest, index voters by $i \in \{1, \ldots, N\}$.  
The number of voters revealed in contest $j$ is given as the sum across voters, $\sum_{i=1}^N r_{ij}.$ 
In Table \ref{tab:counts_revealed}, we compute $N - \sum_{i=1}^N\prod^{K_i}_{j=1}(1 - r_{ij})$, i.e., everyone \emph{except} voters for whom none of their $K_i$ contests are revealed.

We also analyze contest-specific revelation.  
Figure \ref{fig:hist_unanim} studies the mean of $r_{ij}$ for each contest $j$. For generalizability, we omit federal-only ballots, since those few ballots are both substantially more likely to lead to vote revelation and by definition such revelation would reveal a vote for federal office. We also drop three school district contests with few eligible voters for comparability across types of contests.
In Figure \ref{fig:radius}, we limit our attention to the Presidential contest.

\subsubsection{Differential Privacy}

We finally formalize our claim in the Discussion that it is impossible to implement a differential privacy algorithm that will preserve the fidelity of the final vote count in an election determined by one vote. 

Consider the statistic $s$ (e.g. the winner of the election) produced from a private dataset of voters $D$.
The (usually random noise-inducing) processing is denoted $M$, and the value of the statistic is $m$.
The privacy literature considers $M$ to be differentially private if the ratio of $\Pr(M(s, D) = m)$ to $\Pr(M(s, D^\prime) = m)$ is below a user-specified constant $\exp(\epsilon)$, where $D^\prime$ is a dataset that differs from $D$ by the inclusion of one additional row.
Now consider an election is decided by one vote, $D$ is the data set that includes every voter and $D^\prime$ is the data where one voter is excluded. Then if $M$ has no noise, $M(s, D^\prime)$ will report the same winner as $M(s, D)$ with probability 0. Therefore, the ratio of $\Pr(M(s, D) = m)$ to $\Pr(M(s, D^\prime) = m)$ will be infinity, and no value of $\epsilon$ would satisfy the DP criterion.  For $\epsilon$ to approach 0, the result of the election would have to be noised enough that the outcomes of the closest elections would be determined by lot and the presence or absence of any one ballot could never matter. Therefore, differential privacy seems incompatible with the integrity of a single vote. 

Another conceptual challenge is that classic differential privacy is defined on multiple hypothetical versions of the same dataset, comparing whether one person is in the dataset or not. 
This idea may work for a survey sample, where there may be multiple samples from the same population, or for a policy of continuously releasing statistics from a changing population \parencite{wood2018differential}. 
However, an election only happens once. In fact, elections are defined by their {finality}. It is therefore difficult to rely on a notion of multiple versions of the same election.

\newpage
\singlespacing
\setlength\bibitemsep{0.5\baselineskip}
\printbibliography

\newpage

\paragraph{Acknowledgments:} 
For valuable comments and suggestions, we thank participants at the Midwest Political Science Association conference (2023), the Election Science, Reform, and Administration conference (2023), the Conference on Empirical Legal Studies (2024), the University of Pennsylvania Law School's Ad Hoc workshop (2024), the Washington University School of Law Election Law Conference (2024), the Harvard Law School Law and Politics Workshop (2024), and the American Political Science Association conference (2024) as well as members of the State Audit Working Group, members of the North Carolina Board of Elections, and Samuel Baltz, Harvie Branscomb, Paul Burke, Georgie Evans, Adam Friedman, Dave Hoffman, Scott Jarrett, Chris Jerdonek, Christopher T. Kenny, Ray Lutz, Yimeng Li, Chris Warshaw, Benny White, Abby Wood, and Qixuan Yang.
This research was deemed not human subjects research by the IRBs at Yale University and the University of California Los Angeles. A previous draft of the paper was titled ``The Still Secret Ballot: The Limited Privacy Cost of Transparent Election Results.'' 
\paragraph{Funding:} 
The authors received no funding for this research. 

\paragraph{Author Contributions: }
SK, JBL, and MM all contributed to design, data analysis, visualization, writing, and editing. 

\paragraph{Competing Interests: }
The authors have no competing interests.

\paragraph{Data and Materials Availability:} All data needed to evaluate the conclusions in the paper are present in the Supplementary Data Repository at \url{https://doi.org/10.7910/DVN/V83DN5} (Harvard Dataverse).  %

%

%
%
\clearpage
\newgeometry{margin=1.3in}
\appendix
\renewcommand*{\thepage}{SM \arabic{page}}
\setcounter{page}{1}

\begin{center}
{\LARGE \textbf{Supplementary Materials}}

\smallskip

{(Kuriwaki, Lewis, and Morse)}
\end{center}

\setcounter{section}{0}
\setcounter{subsection}{1}
\setcounter{table}{0}
\setcounter{figure}{0}
\setcounter{equation}{0}
\renewcommand{\thefigure}{S\arabic{figure}}
\renewcommand{\thetable}{S\arabic{table}}
\renewcommand{\theequation}{S\arabic{equation}}
\singlespacing

\section{Statistical Model of Vote Revelation \label{sec:model-more}}

In this section, we provide a more general characterization of $\mathrm{E}[R],$ the expected number of publicly revealed votes in a reporting unit. 

\paragraph{Public Revelation}

\begin{figure}[htb]
\caption{\textbf{Determinants of Revelation.} Expected revelations according to our statistical model. The horizontal axis shows the number of voters.  In the left panel, we show simulations from parameters that are conducive to revelation; the right panel shows the opposite. The main text includes parameter details.   \label{fig:model_fig}}
\includegraphics[width=\linewidth]{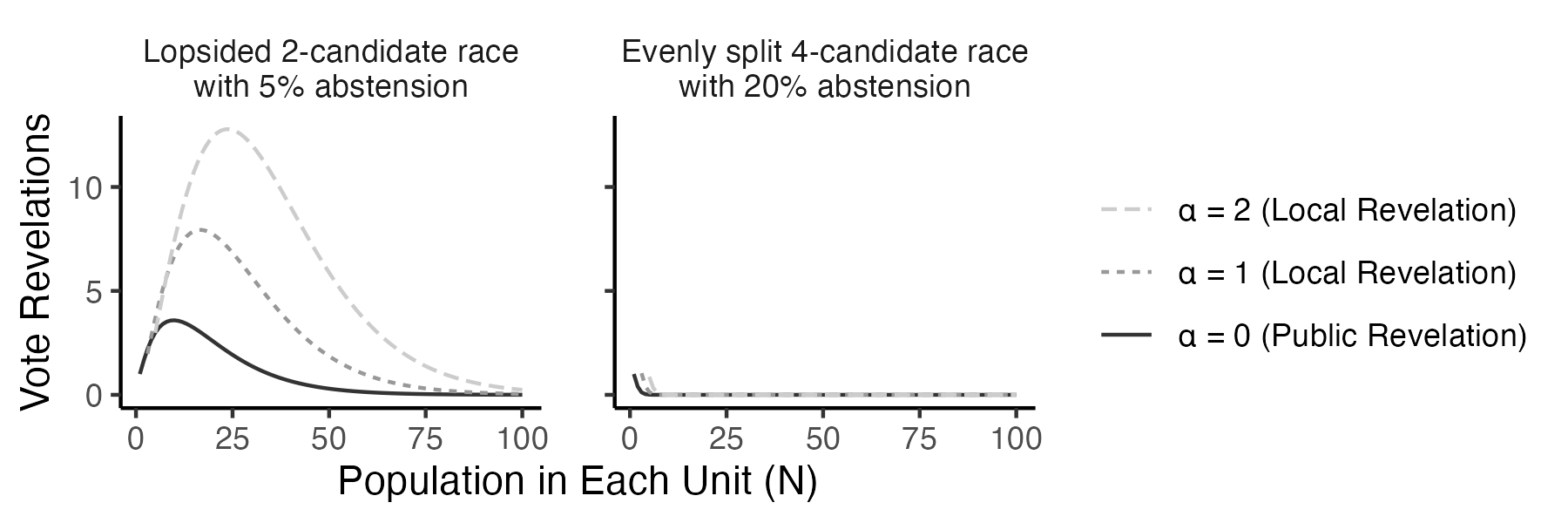}
\end{figure}

Let $s$ be the \emph{a priori} probability that a randomly selected voter in the reporting unit abstains (undervotes or overvotes). Further, allow for more than $H =2$ candidates. 
We slightly re-define $w_h$ to be the probability that a randomly selected voter in the reporting unit will support candidate $h \in \{1,\dots, H\}$ \emph{conditional} on not abstaining.  Then the expected number of revelations is given by:
\begin{align}
\mathrm{E}[R] = N\left(s^N + (1-s)^N\sum_{h=1}^{H} w_h^N \right).
\end{align}
In the special case when $s = 0$, this reduces to $\mathrm{E}[R] = N\left(\sum_{h=1}^{H} w_h^N \right).$ 
This quantity is small when $s$ is large and the vector of $w$ is equally distributed across $H$ candidates. 
While the expected number of revelations at first increases in $N$ in some scenarios, it eventually decreases in $N$ in all scenarios (Figure \ref{fig:model_fig}). 

\paragraph{Local Revelation}
We next consider local revelation at different values of the parameter $\alpha > 0$, where \(\alpha\) denotes the number of otherwise known vote choices. We denote this quantity as $R^\alpha$ to distinguish it from public revelation. 

In the model, the probability of revelation can be calculated from the distribution function for the maximum frequency of a multinomial distribution with parameters
\(\pi = (s, (1-s)w_i, (1-s)w_2, ..., (1-s)w_H)\) and number of trials \(N\).  Revelation occurs if the largest frequency is at least \(N - \alpha\). For example, let $N = 100$, $\alpha = 1$, and $H = 5.$ 
There is $\alpha = 1$ local revelation when at least $100 - 1 = 99$ voters choose the same alternative (candidate or abstain), another alternative gets 1 vote, and the three other alternatives get 0 votes. 
Thus, we are interested in the probability that the maximum number of votes for a candidate across the alternatives is at least 99. This is a well-studied quantity in probability theory. When $\alpha < N/2$,  it is given by  
\begin{align}
\begin{aligned}
\mathrm{E}[R^\alpha] = (N-\alpha)\Bigg[&\underbrace{\sum_{m=N-\alpha}^N {N \choose m} s^m (1-s)^{N-m}}_{\text{Prob. of}~N-\alpha~\text{or more abstentions}}\\
&+ \underbrace{\sum_{m=N-\alpha}^{N} \sum_{h = 1}^H {N \choose m} ((1-s)w_h)^m \left(1-(1-s)w_h\right)^{N-m}}_{\text{Prob. that some candidate received more than}~N-\alpha~\text{votes}}\Bigg].
\end{aligned}
\end{align}
where ${N \choose m}$ indicates the combinations for formula (binomial coefficient) $N$-choose-$m.$ 
In the special case when $s = 0$, then the quantity reduces to 
$$\mathrm{E}[R^\alpha] = (N - \alpha)\left(\sum_{m=N-\alpha}^{N} \sum_{h = 1}^H {N \choose m} w_h^m \left(1-w_h\right)^{N-m} \right).$$ 
All else equal, revelations are smaller when $N$ is large. Revelations increase as $\alpha$ grows. Figure \ref{fig:model_fig} shows how revelations differ when $\alpha = 1$ and $\alpha = 2$.
Note that we multiply the probability of revelation occurring by $N - \alpha$ to get the expected number of revelations because
$\alpha$ of the $N$ votes are not revealed by the election data, but rather by the voters themselves.  The election data can only reveal the remaining $N - \alpha$ votes. 

When $\alpha \geq N/2,$ there is no compact expression of the number of revelations. The reason is that as long as $\alpha < \frac{N}{2}$, it is not possible for two or more alternatives to receive a number 
votes that allow for revelation at the same time.  When this condition holds (which it always does when $\alpha = 0$) then we can find the chances of revelation by summing up the (disjoint) probabilities that each candidate would receive $N-\alpha$ or more votes. However, for larger $\alpha,$ the total probability of revelation can no longer be expressed as a simple sum of the probabilities that each alternative will receive the number of votes required for revelation. 

\paragraph{Illustrative Examples} The solid line in Figure \ref{fig:model_fig} shows how the expected number of revelations changes as $N$ increases for different values of $\alpha$ in each of two different electoral contexts. In the left panel, we consider a lopsided contest with two candidates, $w_1 = 0.95,$ $w_2 = 0.05$, and low abstention, $s = 0.05$. In the right plot, we consider a contest in which support is expected to be evenly split race across four candidates, $w_1 = w_2 = w_3 = w_4 = 0.25,$ with higher abstention, $s = 0.20.$
For a given $N$, the left plot with low abstention and a lopsided contest shows a larger expected number of revelations than the right plot with high abstention and a large field of candidates. 
Also notice that in every scenario considered in the Figure, the expected number of revelations is increasing in $\alpha$, all else equal.

\paragraph{Connection to Empirics}
Our theoretical model illuminates the main components of vote revelation. 
We do not, however, recommend applying our model to predict actual vote revelation.
There are two difficulties in applying our model to predict the magnitude of revelation in a given reporting regime. 
First, the vote probabilities $s$ and $w_h$ are not known \emph{ex ante}.
It may be easy to predict the aggregate vote share for a candidate, but note that these values are defined at the reporting unit level in our model, rather than statewide. 
Second, predicting the implications of increasing $\alpha$ in a real-world setting faces the challenge that $N$ also varies by reporting unit. As our summary statistics show, some granular reporting units have only 1 or 2 voters, while some have thousands. 
For these reasons, the thrust of our study focuses on computing potential revelation from actual data.

\FloatBarrier 
\cleardoublepage

\section{Data Construction}

\subsection{Cast Vote Records} \label{sec:cvr-acquire}

\begin{figure}[tbp]
        \caption{\textbf{Examples of Ballot Images and Cast Vote Records}. (a) An image of an actual ballot from Atkenson et al. (2023). This particular ballot is a rare example in which the voter mistakenly signed their name on the ballot (redacted at the top of the document with a black rectangle).  (b) A mock example of a cast vote record of the same ballot. Information about the precinct, method, and vote choices are stored, but not marks such as the signature.  The precinct and the ballot identifiers are hypothetical values. \label{fig:leon}}
     \centering
     \begin{adjustwidth}{-2cm}{-2cm}
     \begin{subfigure}[t]{0.55\linewidth}
         \centering
         \caption{Ballot Image \label{fig:leon-image}}
         \includegraphics[width=\textwidth]{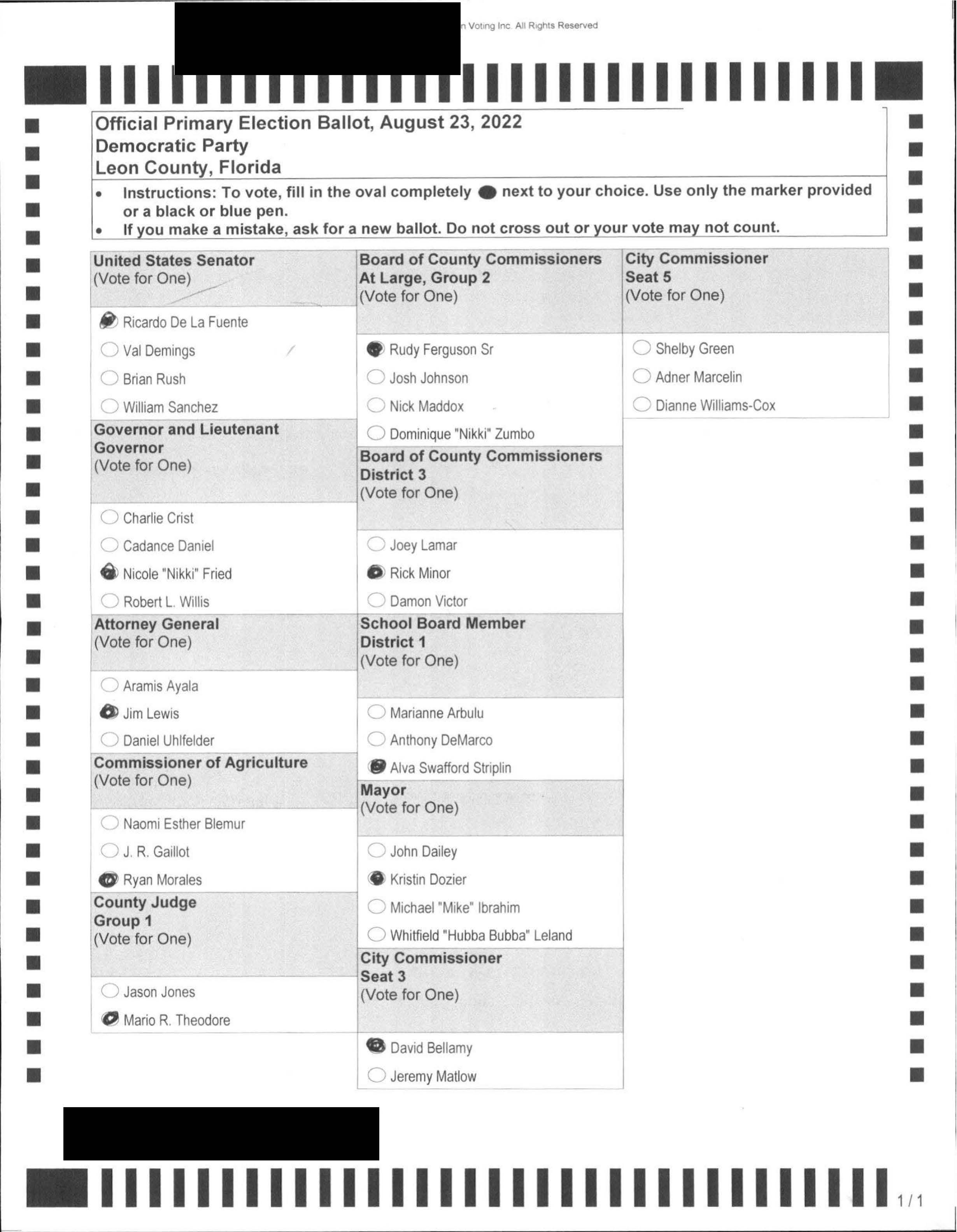}
     \end{subfigure}
     \hfill
     \begin{subfigure}[t]{0.44\linewidth}
         \centering
         \caption{Cast Vote Record representation \label{fig:leon-cvr}}
         \includegraphics[width=\linewidth]{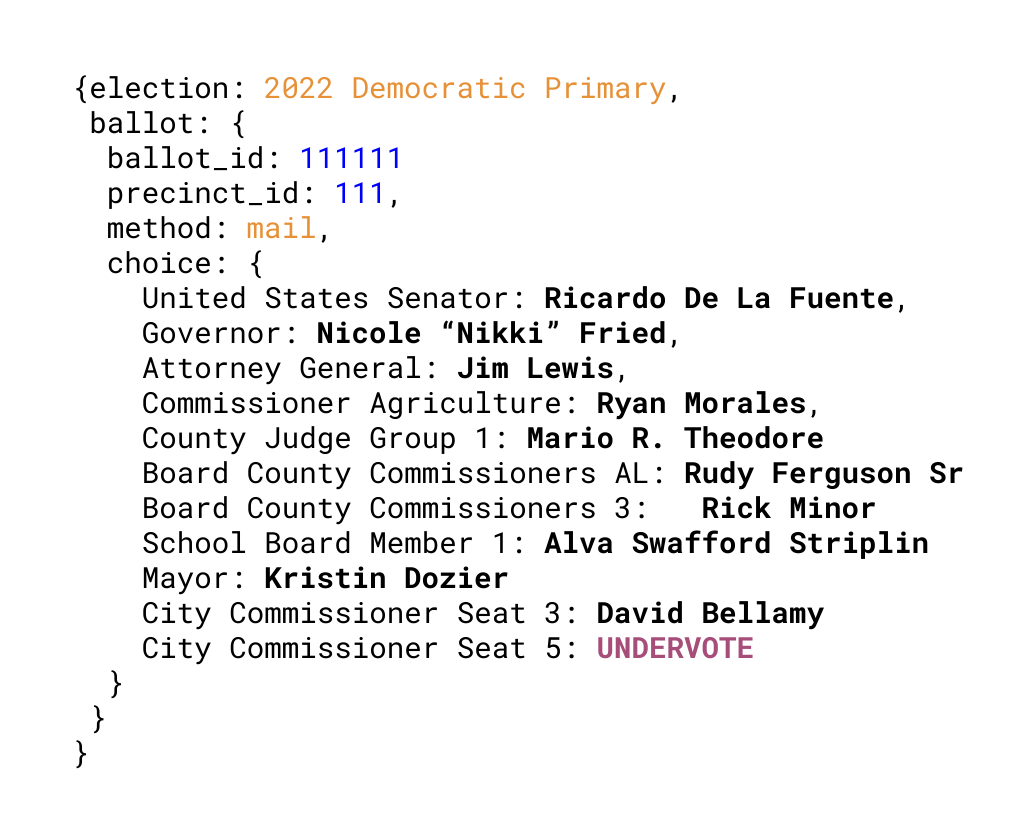}
     \end{subfigure}
     \end{adjustwidth}
\end{figure}

Figure \ref{fig:leon} shows an example of an actual ballot image and its cast vote record representation. 
This particular ballot is a rare example in which the voter mistakenly signed their name on the ballot (redacted at the top of the document with a black rectangle). Through an exhaustive search, Atkenson et al. (2023) find that 22 out of the 118,216 ballots  (0.02\%) cast in Leon County, Florida's 2022 general election include a signature that may or may not indicate the voter's name. 
The code that indicates the precinct is also redacted (near the bottom of the ballot) because of the state's rules to redact precinct identifiers from units where at least one vote method has fewer than 30 votes (Fla. Stat. \S 98.0981(3)(a)1. (2023)).
\FloatBarrier

\begin{figure}[tbh]
\caption{\textbf{Precincts and Ballot Styles}. The figure depicts our definition of precincts and ballot styles in an example of an area split by two district lines. \label{fig:balstyle_def}}
\centering
\includegraphics[width=0.8\linewidth]{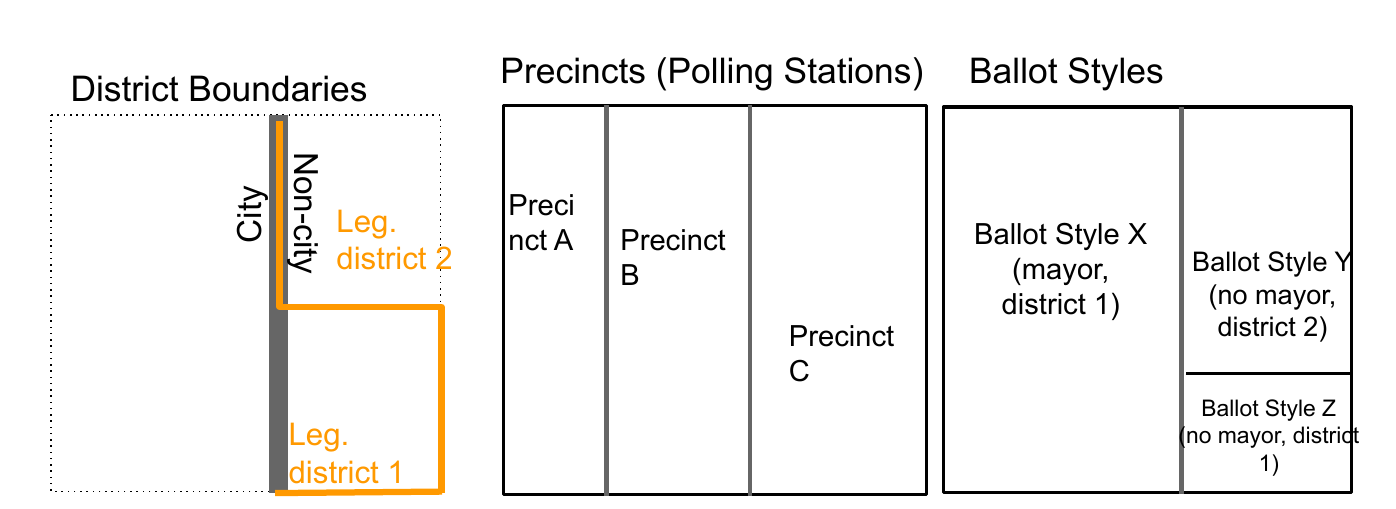}
\end{figure}

Figure \ref{fig:balstyle_def} illustrates the distinction between a precinct and a ballot style.
The figure depicts two political jurisdictions, a city and a legislative district, which are not coterminous.
While all city residents live in the first legislative district, suburban residents are split between two different legislative districts.
The result illustrates the complicated geography of election administration in the United States. In the city, the two precincts will both share the same ballot style (precincts A and B sharing ballot style X), since everyone in the city is in the same municipal and legislative district.
In contrast, in the suburbs, a single precinct will be split by two ballot styles (ballot styles Y and Z in precinct C), since some suburban residents reside in the first legislative district while others reside in the second. 
The result is three distinct ballot styles.

\subsection{Voter File and Voted File Data} \label{sec:voterfile}

Revealing votes requires not only information about the ballots cast, but also about the individuals who cast those ballots.  
In addition to studying what \emph{would} be revealed if cast vote records or granular election returns are combined with a complete and accurate enumeration of the voters casting ballots in the election, we also investigate the {actual} availability of such an enumeration. 

We collect two types of individual-level turnout records, both based on administrative data on voter registration and participation: a commercial voter file produced by L2, a private data vendor, and an administrative file called the ``voted file'' produced by Maricopa County.

The L2 voter file identifies a voter's name, birthdate, and address, as well as whether or not each voter turned out in a specific election and by what voting method. However, consistent with the churn in the electorate, the January 2021 L2 voter file contains 26,819 fewer voters than the votes cast in the November 2020 elections (about 1.3\% of the total votes actually cast).

The voted file, technically called the VM55 file in Maricopa, is a complete record of the voters in each election. It includes precinct and a coarse measure of vote method. Although a voter's ballot style is not included in the voted file, it is possible to map each voter's known registration address to a ballot style with publicly available precinct maps. The only exception to this is if the ballot style in question is a federal-only ballot.
Using the 2020 VM55, we verified that the voted file from Maricopa county matches the number of ballots obtained in the cast vote records, with the exception of the address confidentiality program discuss in the main text.

\subsection{Revelation Agreement by Distance}
\label{sec:radius-appendix}

\begin{figure}[tbh]
\caption{\textbf{Defining Voter's Neighbors}. The blue circle shows a 10-mile radius circle around the center of Maricopa's Sky Hawk precinct.  Sky Hawk precinct, shown in red, cast all 31 of its in-person votes in the 2020 Presidential election for Donald J.\ Trump. To estimate the rate at which voters living within 10 miles of the center of Sky Hawk precinct also supported Trump, we identify those precincts with centroids that are within 10 miles of Sky Hawk's center. Those precincts are shaded blue in the figure. We then calculate Trump support across those precincts.}
\label{fig:explain-radius}
\centering
\includegraphics[width=0.5\linewidth]{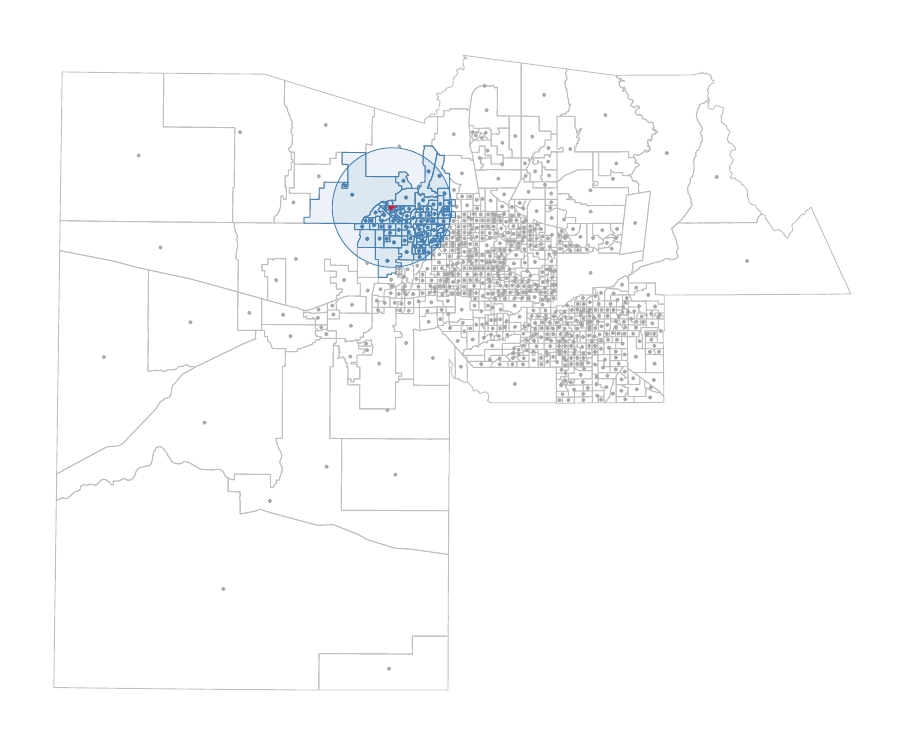}
\end{figure}

This section describes some details of constructing the data underlying Figure \ref{fig:radius}. 
Figure \ref{fig:explain-radius} shows one example of the procedure described in the main text. The blue circle indicates a 10 mile radius around a precinct in which all in-person votes for president were cast for Trump. Figure \ref{fig:radius} shows that the agreement value at $x = 10$ in the left panel is about 65 percent. That means that, among voters whose precinct's centroids are in the blue region of Figure \ref{fig:explain-radius}, about 65 percent of them are Trump voters.

\FloatBarrier

\subsection{Cast Vote Records in Other Counties}
\label{sec:snyder-collection}

To assess the external validity of our results, we turn to the cast vote records from 362 counties curated and published by Kuriwaki, Reece et al. (2024). 
These authors started from a database of cast vote records from the 2020 general election collected by a group of constituents and election skeptics, and released the records for President, U.S. Congress, and State Legislature in counties where the totals implied by the cast vote records sufficiently matched with certified election results.
Conevska et al. (2024) further analyze local offices and ballot measures in these set of counties. 

In the current analysis, we start from the database of Conevska et al. 
 and narrowed down our dataset to counties where the records included sufficient quasi-identifiers to identify the precinct and vote method. 
This left us with %
160
\unskip \ counties that we analyze in Table \ref{tab:snyder-stats}.

\subsection{National Precinct-Level Election Data}
\label{sec:medsl-construction}

We also analyze precinct-level election results from all available states, 
which Baltz et al. (2022) extensively standardized and validated into what we refer to as the MEDSL data.

We used the following procedure to estimate the total number of voters by jurisdiction whose Presidential vote would be revealed if results were reported at the precinct level. In general, computing vote revelation requires knowing the number of residual votes (i.e., undervotes or overvotes) in each reporting unit. 
Results without residual votes are problematic because a reporting unit that appears unanimous based on votes for candidates may not be actually unanimous once residual votes are accounted for.
However, the reporting of residual votes varies state by state. In our analysis of MEDSL data, only 488 out of 2961 counties and parishes reported election results that include the number of residual votes cast by contest by precinct. 

The main idea of our modeling is as follows. We focus on precincts that appear unanimous in jurisdictions that do report residual votes and consider if they are actually unanimous when accounting for residual votes. We fit a logistic regression where the outcome is 1 if the apparently unanimous precinct also had 0 write-in and residual votes (i.e., in fact unanimous), and 0 otherwise (i.e., some voter also voted write-in, undervoted, or both).  Using the regression, we then generate predicted probabilities that each apparently unanimous precinct in the nationwide data is actually unanimous. 

\begin{enumerate}[label=(\alph*)]
\item Starting from MEDSL data, we first consider states that report a ``total votes cast'' by precinct but not residual votes in the Presidential contest, and back out the number of residual votes in a precinct by subtracting the reported votes from the total cast votes. \label{li:national}
\item Subset the data to jurisdictions that report residual votes (including those imputed above) and write-in votes. This limits the data to precincts in %
487
\unskip \ counties. 
\item Subset the data further to precincts in these jurisdictions that are \emph{3-party unanimous}: precincts in which  \emph{only one} of the three 2020 Presidential candidates -- Biden, Trump, and Jorgensen -- have earned votes.  This limits the data to 173 precincts in 46 counties (in the following states: AL, FL, IL, NY, OR, TX, WY).  About %
50
\unskip \ percent of these precincts are Trump precincts in which Trump won votes and Biden won none. Figure \ref{fig:unanim_by_st} shows the voters in such 3-party unanimous precincts as a fraction of the total votes in the entire state.
\item Fit a logistic regression in this data, where the outcome is 1 if the precinct also had 0 write-in and residual votes (i.e., in fact unanimous), and 0 otherwise (i.e., some voter also voted write-in, undervoted, or both). The predictors in the regression are the log of the total votes for the three candidates, and whether the precinct is a Trump district. About %
96
\unskip \ percent of the precincts in the data have an outcome of 1. The logit coefficients are presented in Table \ref{tab:logit_MEDSL}. \label{li:logitrevel}
\item We then return to the national data from \ref{li:national}, and subset it to 3-party unanimous precincts. We therefore take data regardless of whether it reports residual votes by precinct. This results in %
1740
\unskip \ precincts. We then generate predicted probabilities from \ref{li:logitrevel} for each precinct, generating the probability that that precinct is actually revealed. \label{li:pred_revel}
\end{enumerate}

We take the weighted sum of each precinct's vote count in \ref{li:pred_revel}, weighted by the predicted probabilities. This sum generates 11,139, or %
0.0071
\unskip \% of the total number of votes reported in the MEDSL data.

\FloatBarrier

\cleardoublepage

\section{Additional Results}

\FloatBarrier

\subsection{Prevalence of Revelation}

Table \ref{tab:revealed_2022pri} shows the prevalence of revelation in the 2022 primary election. 
Similar to Table \ref{tab:counts_revealed}, it reports public, local, and probabilistic revelation for at least one contest by level of aggregation.
For example, at the precinct level, 68 primary voters have their vote revealed in at least one primary contest, compared to 19 in the general election.

\begin{figure}[htbp]
\centering
\caption{\textbf{Voting Patterns by Contest}. Each point is a contest in the Maricopa 2020 November election, characterized by the level of undervoting (no valid vote for any particular choice) on the horizontal axis and the level of lopsidedness on the vertical axis. The figure only uses contests with two official choices and where the voter makes only one choice.  \label{fig:undervotes-lopsided}}
\includegraphics[width=0.8\linewidth]{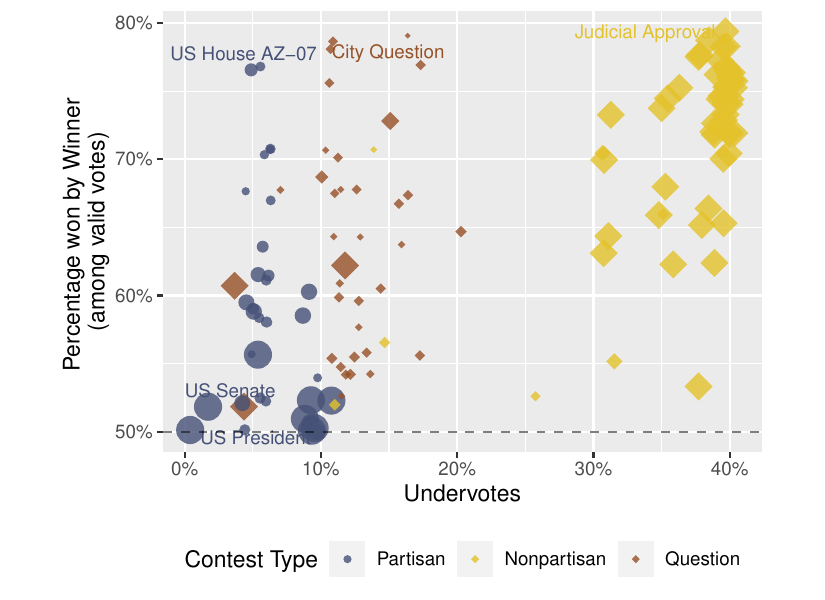}
\end{figure}

Figure \ref{fig:undervotes-lopsided} shows summary statistics of the contests used in Table \ref{tab:counts_revealed} and Figure \ref{fig:hist_unanim}. 
The vote for President was one of the most competitive races countywide with the least amount of undervotes, while nonpartisan judicial retention are lopsided but also feature high amounts of undervoting.

\subsection{Case Study in Context}

\paragraph{Cast Vote Records in Other Counties}

Table \ref{tab:snyder-stats} computes the summary statistics of reporting unit sizes from an alternative set of counties explained in Section \ref{sec:snyder-collection}. We compare the numbers here with Table \ref{tab:data_maricopa} in the main text. 

Table \ref{tab:snyder-demographics} compares the demographic characteristics of the population in these counties using 2020 decennial Census data and the results of the 2020 Presidential election.

\begin{figure}[tbh]
\centering
\caption{\textbf{47-State Analysis of Unanimous Precincts}.
 Each number is the proportion of voters in precincts in which \emph{only one} of the three 2020 Presidential candidates (either Biden, Trump, or Jorgensen) have earned votes.
 This is not equivalent to precincts with vote revelation, because these precincts may also have write-in votes, undervotes, and overvotes.
 New Mexico and Nevada are excluded because they mask the election results in small precincts, and Indiana is excluded because some of its counties do not report precinct results.
 \label{fig:unanim_by_st}
}
\includegraphics[width=0.5\linewidth]{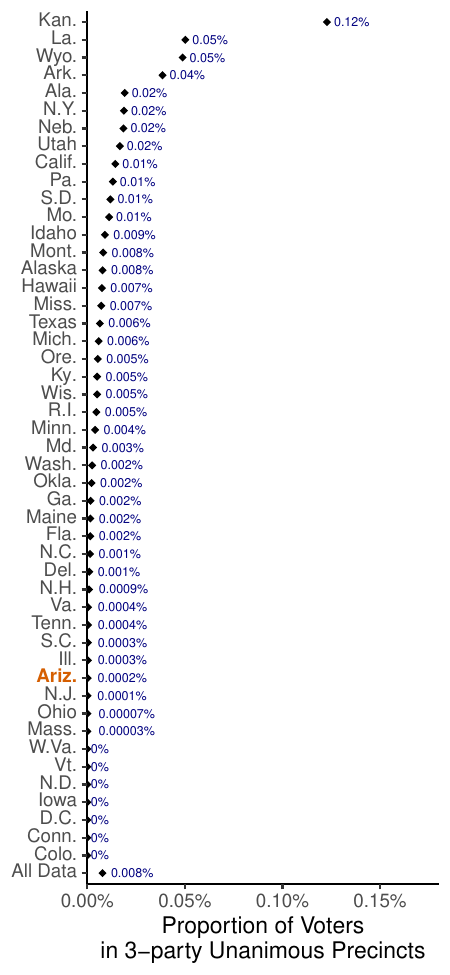}
\end{figure}

\paragraph{National Precinct-Level Results}
Figure \ref{fig:unanim_by_st} reports the percentage of precincts that appear unanimous based on the vote for the three presidential candidates (Biden, Trump, and Jorgensen). 
We call these \emph{three-party unanimous} precincts.
The percent of three-party-unanimous precincts are upper-bounds for the extent of potential vote revelation, because they ignore potential residual votes. 
Any residual vote in a precinct would result in non-unanimity and thus prevent public vote revelation. 
 
\cleardoublepage

\section{Figures and Tables}

\begin{figure}[htb]
\caption{\textbf{Modelling Expected Revelations}. Public vote revelations as a function of the size of a reporting unit. This example depicts a two-candidate contest without abstention where the probability of voting for the leading candidate is 0.7. \label{fig:model}}
\centering
\includegraphics[width=0.45\textwidth]{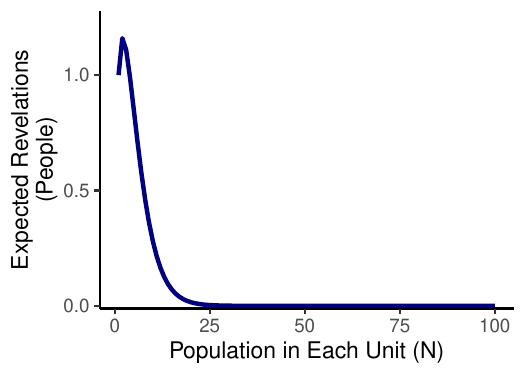}
\end{figure}

\begin{table}[tbh]
\centering
\caption{\textbf{Revelation in the 2022 Primary Elections}. Follows Table \ref{tab:counts_revealed} but shows the 2022 Primary Election rather than the 2020 General Election. \label{tab:revealed_2022pri}}
\includegraphics[width=0.8\linewidth]{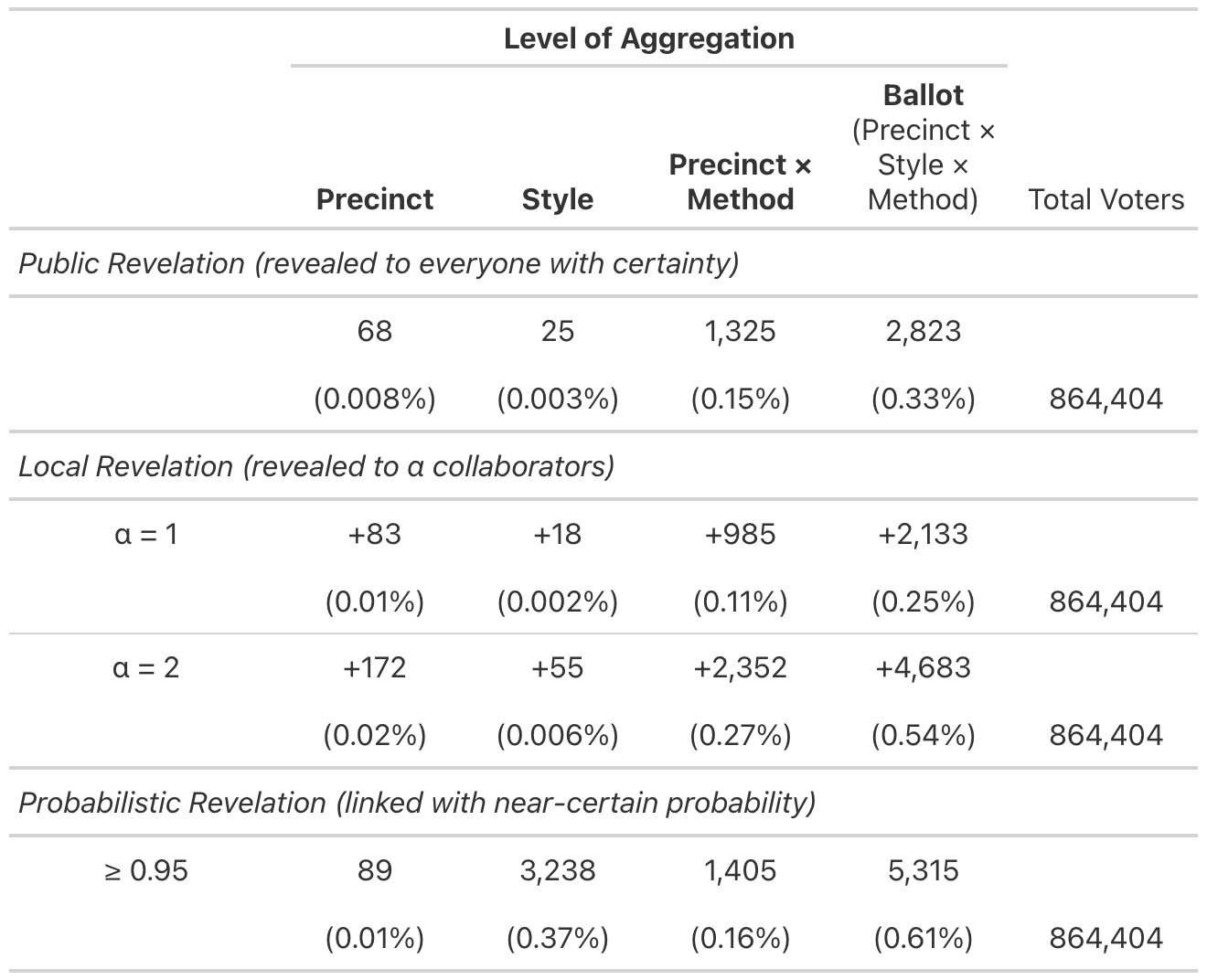}
\end{table}

\begin{table}[tbh]
\caption{\textbf{Summary Statistics for Comparison Dataset} \label{tab:snyder-stats}}
\begin{tabular}{rcccccc}
\toprule
 &  & \multicolumn{5}{c}{Voters per 100,000 in units of size:} \\ 
\cmidrule(lr){3-7}
Reporting Unit & Unique units & 1 & $\le 2$ & $\le 5$ & $\le 10$ & $\le 30$\\ 
\midrule\addlinespace[2.5pt]
Precinct & $8,001$ & $0$ & $1$ & $5$ & $12$ & $49$ \\ 
Ballot Style & $4,498$ & $1$ & $2$ & $6$ & $13$ & $46$ \\ 
Precinct $\times$ Ballot Style & $12,465$ & $3$ & $7$ & $16$ & $34$ & $117$ \\ 
Ballot Equivalent & $28,004$ & $10$ & $23$ & $67$ & $158$ & $559$ \\ 
\emph{... only mail method} & $10,173$ & $4$ & $8$ & $20$ & $43$ & $191$ \\ 
\emph{... only in-person method} & $14,787$ & $10$ & $24$ & $71$ & $174$ & $828$\\ 
\emph{... only provisional method} & $3,042$ & $422$ & $983$ & $3,022$ & $7,096$ & $15,991$\\ 
\bottomrule
\end{tabular} \end{table}
\begin{table}
\centering
\caption{\textbf{Demographics of Comparison Dataset} \label{tab:snyder-demographics}}
\begin{tabular}{lccc}
\toprule
Variable & Maricopa & Comparison & All 50 States \\ 
\midrule\addlinespace[2.5pt]
Percent White & 59.8 & 52.2 & 61.2 \\ 
Percent Black & 5.9 & 14.5 & 12.3 \\ 
Percent Hispanic & 30.6 & 23.2 & 19.5 \\ 
Percent Urban & 97.7 & 85.4 & 80.1 \\ 
Percent Biden Voteshare & 51.1 & 55.3 & 52.2 \\\midrule
Population (millions) & 4.4 M & 23.2 M & 334.7 M \\
Counties & 1 & 160 & 3,221\\
\bottomrule
\end{tabular}
\end{table}

\begin{table}[hbp]
\centering
\caption{\textbf{Model predicting unanimous results among precincts where all candidate votes were for one candidate.} Table shows logit coefficients and standard errors in parentheses. Unanimous results are defined here as having voted for a single Presidential candidate with no overvotes or undervotes. The variable ``support for Trump'' distinguishes all-Trump precincts from all-Biden precincts. \label{tab:logit_MEDSL}}
\small
\begin{tabular}[t]{lc}
\toprule
  & Outcome: Actual Unanimity\\
\midrule
Intercept & 3.62\\
 & (0.69)\\
log(Total votes for Biden/Trump/Jorgensen) & -0.41\\
 & (0.25)\\
Support for Trump & 0.27\\
 & (0.79)\\
\midrule
Mean of Outcome & 0.96\\
N & 173\\
\bottomrule
\end{tabular}
\end{table}

\end{document}